\documentclass[10pt,conference]{IEEEtran}

\IEEEoverridecommandlockouts
\usepackage{cite}
\usepackage{amsmath,amssymb,amsfonts}
\usepackage{algorithmic}
\usepackage{graphicx}
\usepackage{textcomp}
\usepackage{xcolor}
\def\BibTeX{{\rm B\kern-.05em{\sc i\kern-.025em b}\kern-.08em
    T\kern-.1667em\lower.7ex\hbox{E}\kern-.125emX}}

\usepackage{graphicx}
\usepackage{algorithmic}
\usepackage{textcomp}
\usepackage{xcolor}
\usepackage{etoolbox}
\usepackage{multirow}
\usepackage{url}
\usepackage{multirow,booktabs}
\usepackage{subcaption}
\usepackage{wrapfig}
\usepackage{float}
\usepackage{float}
\usepackage{tcolorbox}
\usepackage[flushleft]{threeparttable}
\usepackage{relsize}
\usepackage[inline]{enumitem}
\usepackage{amsmath}

\usepackage{algorithm2e}

\restylefloat{figure}

\newcommand{\revise}[1]{{\color{black}{#1}}}

\newcommand{\revisearxiv}[1]{{\color{black}{#1}}}

\newcommand{\ourapp}{\textsc{LEO}}
\newcommand{\ourmechanism}{underlying-data-distribution
mechanism}

\newcommand{\rqone}{Can our method successfully identify OOD source code data? How effective is it compared to baselines?}

\newcommand{\rqtwo}{How does our method perform in different cases of the in-distribution data when it consists of more than one CWE source code data category?}

\newcommand{\rqthree}{Do the innovative cluster-contrastive learning and \ourmechanism~ help improve the code data representation learning for boosting the model performance in OOD source code data identification?}

\begin{document}

\title{Deep Learning-Based Out-of-distribution Source Code Data Identification: How Far Have We Gone?}

\author{Van Nguyen\textsuperscript{*}$^{,1,2}$\thanks{*Corresponding Author (van.nguyen1@monash.edu). Van Nguyen is a Postdoctoral Research Fellow at the Department of Software Systems and Cybersecurity at Monash University, Australia. Additionally, he is an Affiliate at CSIRO’s Data61, Australia.} and Xingliang Yuan$^{1}$ and Tingmin Wu$^{2}$ and Surya Nepal$^{2}$ \\ Marthie Grobler$^{2}$ and Carsten Rudolph$^{1}$ \\ \\
$^{1}$ Monash University, Australia\\
$^{2}$ CSIRO's Data61, Australia}

%Van Nguyen and Xingliang Yuan and Tingmin Wu and Surya Nepal and Marthie Grobler and Carsten Rudolph

\maketitle

\begin{abstract}
Software vulnerabilities (SVs) have become a common, serious, and crucial concern to safety-critical security systems. That leads to significant progress in the use of AI-based methods for software vulnerability detection (SVD). In practice, although AI-based methods have been achieving promising performances in SVD and other domain applications (e.g., computer vision), they are well-known to fail in detecting the ground-truth label of input data (referred to as out-of-distribution, OOD, data) lying far away from the training data distribution (i.e., in-distribution, ID). This drawback leads to serious issues where the models fail to indicate when they are likely mistaken. To address this problem, OOD detectors \textit{(i.e., determining whether an input is ID or OOD)} have been applied before feeding the input data to the downstream AI-based modules.

While OOD detection has been widely designed for computer vision and medical diagnosis applications, automated AI-based techniques for OOD source code data detection have not yet been well-studied and explored. To this end, in this paper, we propose an innovative deep learning-based approach addressing the OOD source code data identification problem. Our method is derived from an information-theoretic perspective with the use of innovative cluster-contrastive learning to effectively learn and leverage source code characteristics, enhancing data representation learning for solving the problem. \textit{The rigorous and comprehensive experiments on real-world source code datasets show the effectiveness and advancement of our approach compared to state-of-the-art baselines by a wide margin}. \revise{In short, on average, our method achieves a significantly higher performance from around 15.27\%, 7.39\%, and 4.93\% on the FPR, AUROC, and AUPR measures, respectively, in comparison with the baselines.}

\end{abstract}

\section{Introduction}\label{sec:introduction}

Software vulnerabilities (SVs), known as specific flaws or oversights in software programs allowing attackers to exploit the code base and potentially undertake dangerous activities (e.g., exposing sensitive information or taking control of a computer system) \cite{Dowd2006}, have become a common, serious, and crucial issue to safety-critical security. There have been many methods proposed for software vulnerability detection (SVD) ranging from open-source to commercial tools, and from manual to automatic methods \cite{Neuhaus:2007:PVS, shin2011evaluating, yamaguchi2011vulnerability, Grieco2016, Li2016:VAV, KimWLO17, VulDeePecker2018, Duan2019, Cheng2019, nguyen2019deep, Zhuang2020, van-dual-dan-2020, van-ijcnn2021, ReGVD2022, Michael2023VulExplainer, Michael2023AIBugHunter} based on machine learning and deep learning approaches (i.e., AI-based models).

In practice, while AI-based models demonstrate notable performances in multiple domain applications such as software vulnerability detection (SVD), autonomous driving, and biometric authentication, they often obtain poor performances or fail to detect the ground-truth labels (e.g., the categories of traffic signs or the vulnerability (e.g., vulnerable or non-vulnerable) of the source code samples) when encountering the input data (commonly referred to as out-of-distribution, OOD, data) significantly lying far away from the training data distribution \cite{goodfellow2015explaining, AmodeiOSCSM16, ReGVD2022,llmsvd2024}. This limitation poses a serious challenge as the models struggle to accurately identify instances when they are likely to make mistakes. In particular, deep learning-based classifiers have been shown to incorrectly classify the ground-truth labels of out-of-distribution data samples with high confidence \cite{goodfellow2015explaining, NguyenYC14}. That can limit their adoption as well as cause unexpected crucial accidents in safety-critical systems such as medical and security domains. \textit{This problem is particularly severe within the cyber security realm}. Hackers can create new vulnerabilities (with the assistance of domain experts and AI-based tools, e.g., generative large language models) and exploit these zero-day (a.k.a. out-of-distribution) vulnerabilities associated with emerging CWE categories\footnote[1]{CWE stands for Common Weakness Enumeration. It is a community-developed list of common security weaknesses that software developers, testers, and security professionals can use to identify and mitigate vulnerabilities in software systems \cite{cwe}.} to compromise software security systems until appropriate measures are taken to address the issue.

The problems of out-of-distribution (a.k.a anomaly) result in rapid progress in the use of advanced machine learning and deep learning for OOD detection before feeding input data samples to the downstream AI-based modules. However, though achieving significantly promising performance for OOD detection, almost all existing methods are mainly based on or applicable to images in computer vision \cite{HendrycksG16c, Hendrycks18, Lee18a, Lee18b, Sehwag21, sun2022outofdistribution} or genomics sequences in medical diagnosis \cite{Ren19}, automated deep learning-based techniques for detecting OOD source code data have not yet been well studied and explored. Furthermore, we observed that these methods cannot be straightforwardly applied in the context of out-of-distribution source code identification due to the differing nature of source code data compared with images and genomics sequences. Motivated by this problem, in this paper, we study the research questions:

\vspace{1mm}
\colorbox{gray!10}{
\begin{minipage}{0.91\columnwidth}
\revise{Given an in-distribution (ID) source code dataset $D_{in}$ (associated with one or multiple CWE categories having both vulnerable and non-vulnerable source code data samples), how to develop an effective method that can successfully identify whether an input source code data sample is from $D_{in}$ or from an out-of-distribution (OOD) CWE category? What are the main challenges and characteristics of source code data that need to be considered? (in the scope of our paper, we name this problem as OOD source code data identification)}
\end{minipage}
}
\vspace{1mm}

When considering source code data, we notice that various instances of vulnerable source code data, associated with different CWE categories (e.g., buffer overflow error or resource management error), frequently stem from distinct hidden vulnerability patterns \cite{cwe}. Furthermore, in each vulnerable source code sample (i.e., a function or a program), the hidden vulnerability pattern emerges from the intricate semantic relationships between vulnerability-relevant source code statements. Therefore, to figure out the difference between source code data from \revise{various CWE categories} that support identifying \revise{OOD CWE source code data, especially those that are vulnerable}, the corresponding method needs to have the ability to learn and recognize the important and vulnerability-relevant source code statements in each source code sample, particularly within the vulnerable data.

In addition, we observed some characteristics of source code data as follows: (i) source code data are complicated and different from texts or images due to consisting of complex semantic and syntactic relationships between code statements and code tokens, (ii) different source code data can share a significant amount of common background information (e.g., non-vulnerable code statements) while there are only a few specific source code statements causing the data to be different and vulnerable. This problem requires an appropriate solution to learn and figure out vulnerable parts inside vulnerable data for being able to distinguish non-vulnerable and vulnerable source code data samples, and (iii) although having the same vulnerability label (e.g., vulnerability type, i.e., buffer overflow error, or vulnerable one in general), there are still many different hidden vulnerability patterns causing source code data samples to be classified as vulnerable. Therefore, an elegant mechanism for automatically figuring out and separating vulnerable source code samples consisting of different hidden vulnerability patterns is essential for ensuring robust representation learning for OOD source code data identification.

To this end, in this paper, we propose a novel deep learning-based approach that can learn and leverage the characteristics of source code data to automatically figure out the important and vulnerability-relevant source code statements, forming the corresponding hidden vulnerability patterns to vulnerable source code data. In addition, our approach can also utilize the semantic relationships of vulnerability patterns inside and between source code data samples to boost the data representation learning that facilitates solving the OOD source code data identification problem. Via this paper, we demonstrate that the ability to capture the characteristics of source code data as well as figure out and leverage the relationships of hidden vulnerability patterns inside and between source code data samples for boosting the data representation learning plays a key role in OOD source code data identification.

In particular, our approach consists of two phases. \textbf{In the training phase}, we propose an elegant training principle based on information theory and cluster-contrastive learning for effectively learning and leveraging the characteristics of source code data to boost the data representation learning process.
During the training phase, our proposed approach enhances the learning of semantic relationships inside and between source code data samples for forming clear boundaries between source code data consisting of different hidden vulnerability patterns (i.e., note that in the latent space, the representations of vulnerable source code data sharing the same hidden vulnerability pattern tend to form a corresponding cluster) in the latent space for the training set of the in-distribution data ($D_{in}$). \textbf{In the testing (inference) phase}, inspired by \cite{Lee18b, Winkens2020, Sehwag21}, 
we use the cluster-conditioned detection method (using the Mahalanobis distance \cite{Mahalanobis}) to calculate the outlier score $s(X)$ for each testing input data sample $X$ in the latent space with respect to the training data set (i.e., $D_{in}$). Based on the outlier score $s(X)$, we determine whether $X$ is from out-of-distribution. Please refer to Section \ref{sec:framework} for a comprehensive explanation, visualization, and algorithm of our method.

\vspace{0mm}
In summary, our key contributions are as follows:
\begin{itemize}
\vspace{-2mm}
\item We study an important problem of identifying OOD source code data, which plays a crucial role in strengthening the defense of software security systems. Automated machine learning and deep learning-based techniques for this problem have not yet been well studied and explored. 
\vspace{-3mm}
\item \revise{We propose an innovative information-theoretic learning-based approach to identify out-of-distribution source code data samples (i.e., functions or programs). Our approach can automatically learn the characteristics of source code data as well as also figure out and leverage the semantic relationships of hidden vulnerability patterns inside and between source code data samples to boost the data representation learning, enabling the identification of OOD source code data.} To the best of our knowledge, our work is one of the first methods proposed to solve the problem. It can serve as an effective baseline for solving the OOD source code data identification problem.
\vspace{1mm}
\item We comprehensively evaluate and compare our proposed approach with state-of-the-art baselines on real-world source code data covering many CWE categories. Our extensive experiments show that our approach obtains a significantly higher performance than the baselines in three main measures including FPR (at TPR 95\%), AUROC, and AUPR, widely utilized in out-of-distribution identification methods, as noted in \cite{Hendrycks18, Ren19, Sehwag21}.

\end{itemize}
\section{Motivations}\label{sec:motivations}

\subsection{\textbf{The impact of out-of-distribution (OOD) source code data}}

\revise{Out-of-distribution (a.k.a. zero-day) source code data, especially vulnerable ones, are considered highly serious in the realm of cyber-security due to their potential for exploitation before the affected software (system) has a chance to defend against them. Attackers can exploit OOD vulnerabilities (associated with emerging CWE categories) to compromise systems before the developers or security community becomes aware. This gives hackers a head start in deploying attacks.

The effect of out-of-distribution vulnerability attacks can be far-reaching, as they can compromise sensitive data, disrupt critical infrastructure, and cause financial losses. For instance, the WannaCry ransomware attack in 2017 exploited a vulnerability in Microsoft Windows and affected more than 200,000 computers in 150 countries. Also in 2017, a zero-day vulnerability in the Equifax credit reporting agency's software was exploited, resulting in the theft of sensitive personal and financial information belonging to over 140 million customers. Consequently, Equifax suffered substantial financial and reputational harm, paying over \$1 billion in compensation, facing lawsuits, and dealing with regulatory penalties.}

\subsection{\textbf{The need for OOD source code data identification}}

\revise{The incidents resulting from out-of-distribution (OOD) vulnerability attacks have grown increasingly severe, particularly in the era of generative machine learning. With the assistance of domain experts and AI-based tools, e.g., generative large language models, hackers can create and exploit unknown vulnerabilities to compromise software security systems until appropriate measures are taken to address the issue.

Software vulnerability detection (SVD) AI-based methods have been proposed and deployed in software security systems to enhance their defense in detecting the data's vulnerability (e.g., vulnerable or non-vulnerable). Although achieving promising performances for SVD, AI-based methods can only work well with the input data from the same distribution as the data used to train the models. When the data have different representations or lie far away from the training data distribution, AI-based methods often achieve poor performances or fail to detect the data vulnerability \cite{d2a2021,ReGVD2022,llmsvd2024}.

Out-of-distribution vulnerabilities associated with emerging CWE source code categories arise annually \cite{cwe}, prompting the need for solutions to enhance the defense of software security systems. \textit{To address this issue, the identification of OOD source code data appears as a promising approach.} The OOD source code data identification task helps determine if an input is from a learned in-distribution (ID) or out-of-distribution (OOD). That enables the system and security experts to take appropriate precautions and actions. In particular, this stage assists in identifying potential OOD CWE source code data outside the established in-distribution CWE source code data categories to harden the software security systems before the input data is fed into subsequent AI-based modules.}
\section{Related Work}\label{sec:related_work}

\subsection{\textbf{Out-of-distribution (OOD) data identification}}

There have been many AI-based methods (i.e., machine learning-based and deep learning-based approaches) introduced for out-of-distribution detection widely applied to computer vision and medical diagnosis applications \cite{Le2014, nguyen2014, Duong2015, OordKK16, SalimansKCK17, LiangLS17, GuoPSW17, Lakshminarayanan-2016, Lee18a, Hendrycks18, Lee18b, Nalisnick18, Choi18, Kingma18, Ren19, Hendrycks2019, Winkens2020, Sehwag21}. Although existing methods have achieved promising results for out-of-distribution detection, they are mainly based on or only applicable to computer vision applications or medical diagnosis. Automated deep learning-based techniques for out-of-distribution source code data identification have not yet been well studied. \revise{In our paper, we apply some state-of-the-art out-of-distribution detection methods (e.g., \cite{Hendrycks2017dnn, Hendrycks18, Sehwag21}) to the problem of identifying OOD source code data. We then proceed to evaluate and compare the performance of these models against our method (please refer to the Experiment section (Section \ref{sec:experiments}) for details)}.

\vspace{-1mm}
\subsection{\textbf{Software vulnerability detection (SVD)}}
\vspace{-1mm}

AI-based approaches have been widely proposed for SVD, ranging from utilizing handcrafted features manually selected by domain experts \cite{yamaguchi2011vulnerability, shin2011evaluating, Li2016:VAV, Grieco2016, KimWLO17}, to leveraging automatic feature learning through deep learning-based methods \cite{VulDeePecker2018,jun_2018, Dam2018, Li2018SySeVR, Duan2019, Cheng2019, Zhuang2020, ReGVD2022, nguyen2022info, nguyen2022cross, Fu2023, Fu2023quantization, Fu2023repair, Fu2023vulrepair}, showcasing notable advancements in the field. Although achieving significant performances in SVD, AI-based methods often achieve poor performances or struggle when encountering input data situated distantly from the training data distributions \cite{d2a2021,ReGVD2022,llmsvd2024}. \revise{In our paper, we also investigate if effective methods in SVD can be applied to identifying out-of-distribution source code data.} \revise{The rationale is that an effective model should produce high-quality data representations that facilitate distinguishing the data between in-distribution (ID) and out-of-distribution (OOD).} 

We employ effective and state-of-the-art SVD methods including \cite{VulDeePecker2018, ReGVD2022, CodeBERT2020}. VulDeePecker \cite{VulDeePecker2018} is considered one of the most effective deep learning-based SVD methods leveraging the power of bi-directional long-short-term memory (LSTM) networks. CodeBERT \cite{CodeBERT2020} is a pre-trained model (based on the Transformer's encoder framework \cite{transformers-Vaswani17}) that specializes in the programming language. Via CodeBERT, we aim to investigate how large language models work in the context of out-of-distribution source code data identification. ReGVD \cite{ReGVD2022} is a simple yet effective graph neural network-based model for the SVD problem.

\vspace{-1mm}
\subsection{\textbf{The difference tasks between source code vulnerability detection and OOD source code identification}}
\vspace{-1mm}

\revise{The task of identifying OOD source code data differs from that of source code vulnerability detection. In the OOD identification task, the objective is solely to determine whether input data is from established in-distribution categories or from an out-of-distribution category while in vulnerability detection, we aim to detect the vulnerability (e.g., vulnerable or non-vulnerable) of the data.} The OOD identification stage assists the security systems and experts in identifying OOD source code data outside the established in-distribution data for conducting appropriate precautions and actions before feeding input data into subsequent AI-based modules, including the vulnerability detection phase.
\vspace{3mm}
\section{The proposed \ourapp~approach}\label{sec:framework}

\subsection{\textbf{The problem statement}}
We denote $D_{in}$ as a real-world source code dataset (associated with one or multiple CWE categories) consisting of $\{(X_{1}, Y_{1}),\dots,(X_{N_{D_{in}}}, Y_{N_{D_{in}}})\}$ where $X_i$ is a source code data sample (i.e., a function or a program) comprised of $L$ code statements from $\boldsymbol{x}_{1}$ to $\boldsymbol{x}_{L}$ while $X_i$'s vulnerability label $Y_{i}\in\{0,1\}$ (i.e., $0$: non-vulnerable and $1$: vulnerable). From the dataset $D_{in}$, we can build AI-based (machine learning or deep learning) approaches to detect the vulnerability (i.e., benign or vulnerable) of the source code data sampled from the distribution $D_{in}$; however, they are known to fail against the data samples (outliers or out-of-distribution) that lie far away from the data distribution $D_{in}$ with high confidence.

In software security systems, the OOD source code data identification process enables the model and security experts to take appropriate precautions and actions, before feeding source code input data into subsequent AI-based modules. \textbf{In our study, given an in-distribution source code dataset $D_{in}$, we aim to develop a deep learning-based approach that can effectively identify whether an input source code data sample is from a learned in-distribution $D_{in}$ or from an out-of-distribution $D_{out}$ CWE category}.

\vspace{0mm}
\subsection{\textbf{Methodology}}

In what follows,  we present the details of how our \ourapp~method works and addresses \revise{the out-of-distribution source code data identification problem}. We first describe the process of our \ourapp~method taking into account the source code characteristics and working to automatically learn and figure out the important and vulnerability-relevant source code statements, forming the corresponding hidden vulnerability patterns in vulnerable source code data, contributing to robust source code data representation learning for facilitating \revise{out-of-distribution source code data identification}. We then present our effective solution for leveraging the semantic relationships of hidden vulnerability patterns inside and between the source code samples (i.e., functions or programs) by utilizing innovative cluster-contrastive learning to further improve the source code data representation learning process.

\begin{figure}[ht]%
\vspace{0mm}
\begin{centering}
\begin{tabular}{c}
\includegraphics[width=0.43\textwidth]{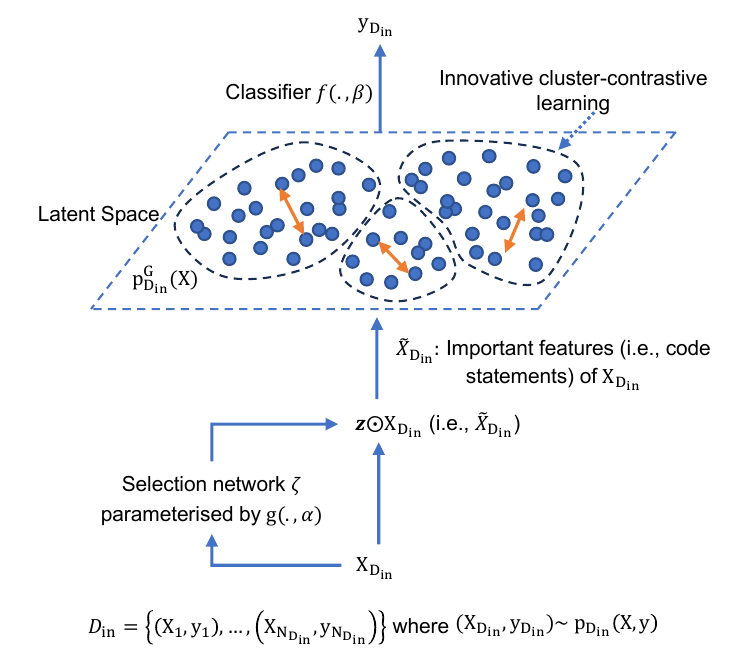}\tabularnewline
\end{tabular}
\par\end{centering}
\vspace{1mm}
\caption{A visualization of our proposed \ourapp~method for effectively improving the source code data representation learning process \revisearxiv{to solve the out-of-distribution (OOD) source code data identification problem.}}\label{fig:deeplearning_v2}
\vspace{-3mm}
\end{figure}%

\vspace{1mm}
\subsubsection{\textbf{The selection of important features (i.e., source code statements)}}\label{sec:selection_network}

As depicted in Figure \ref{fig:deeplearning_v2}, our \ourapp~method starts with a selection network $\zeta$ for automatically learning and figuring out the crucial and \revise{label-relevant code statements} in each source code data (a function or a program). It is important to highlight that to vulnerable source code data, the identified code statements constitute the corresponding vulnerability patterns, rendering the data vulnerable. Through the selection network with the use of innovative cluster-contrastive learning (refer to section \ref{sec:cluster_contrastive_learning} for details), in the latent space, the source code data containing different hidden vulnerability patterns will be trained to have different representations, causing them to be situated far apart from each other while the source code data sharing similar hidden vulnerability patterns are encouraged to remain closely clustered. This process elegantly facilitates \revise{out-of-distribution source code data identification.}

\vspace{1mm}
\paragraph{\textbf{Motivations for important code statement selection to boost the data representation learning process}}

In every source code data sample, a few certain code statements typically hold greater significance than others in contributing to the corresponding vulnerability, especially in vulnerable source code data. These important and vulnerability-relevant statements often constitute the underlying hidden vulnerability patterns within the vulnerable source code data, as illustrated in examples in Figure \ref{fig:boe_update}. In contrast, the vulnerability-irrelevant code statements can be shared among different non-vulnerable and vulnerable data, potentially compromising the model's capability to learn the representations of vulnerable data. \revise{Hence, the automatic learning and identification of critical code statements within each source code data sample play a pivotal role in advancing the data representation learning process. This facilitates the distinction not only between non-vulnerable and vulnerable data but also between data exhibiting different hidden vulnerability patterns.}

\vspace{1mm}
\paragraph{\textbf{Selection network for choosing important and label-relevant code statements in each source code data sample}}

\begin{figure}[th]
\vspace{0mm}
\begin{centering}
\includegraphics[scale=0.63]{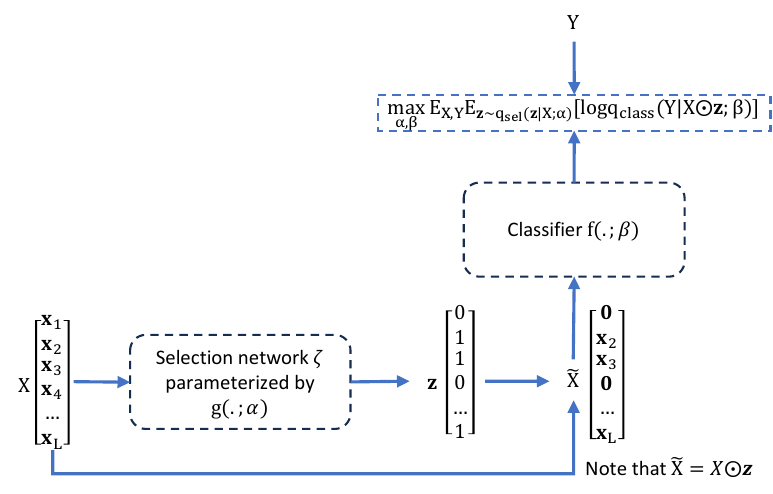}
\par\end{centering}
\vspace{0mm}
\caption{A visualization of the joint training process of the classifier and the selection model process guided by maximizing mutual information between $\tilde{X}$ and $Y$.}\label{fig:visualizationfor_eq_1_2}
\vspace{-3mm}
\end{figure}

\revise{Giving a source code data sample $X$ consisting of $L$ code statements from $\mathbf{x}_{1}$ to $\mathbf{x}_{L}$, to select the important and vulnerability label-relevant code statements (denoted by $\tilde{X}$) of $X$, we design a selection model $\zeta$ parameterized by a neural network $g(.,\alpha)$. The network $g$ takes $X=\{\mathbf{x}_{i}\}_{i=1}^{L}$ as the input and outputs the corresponding $\mathbf{p}=\{p_{i}\}_{i=1}^{L}$ where each $p_{i}$ presents the probability of $\mathbf{x}_{i}$ related to the vulnerability $Y$ of $X$. In practice, each code statement $\mathbf{x}_{i}$ is represented as a vector using a learnable embedding method described in the data processing and embedding section.

We then construct $\tilde{X}=\zeta\left(X\right)$ (i.e., the subset code statements that lead to the vulnerability $Y$ of the function $X$) by $\tilde{X}=X\odot \mathbf{z}$ with $\mathbf{z} \sim \mathrm{MultiBernoulli}(\mathbf{p}) = \prod_{i=1}^{L}\text{Bernoulli}(p_{i})$ and the element-wise product $\odot$ as depicted in Figure \ref{fig:visualizationfor_eq_1_2}. Note that we can view the selection model $\zeta$ as a distribution $q_{sel}(\mathbf{z}|X;\alpha)$ over a selector variable $\mathbf{z}$ which indicates the important features (i.e., source code statements) of a given sample $X$ (i.e., each $z_{i}$ in $\mathbf{z}=\{z_{i}\}_{i=1}^{L}$ indicates if $\mathbf{x}_{i}$ is significant in leading the vulnerability $Y$ of $X$. Specifically, if  $z_{i}$ is equal to $1$, the statement $\mathbf{x}_{i}$ plays an important role in causing the vulnerability $Y$).}

\vspace{0mm}
To make the selection process (i.e., consisting of sampling operations from a Multi-Bernoulli distribution) continuous and differentiable during training, we apply the Gumbel-Softmax trick {\cite{jang2016categorical, MaddisonMT16}} for relaxing each Bernoulli variable $z_{i}$.
\vspace{3mm}
\paragraph{\textbf{Information theory-based learning to guide the selection process}}

\revise{It is important to highlight that the source code data $X$ (as well as its $\tilde{X}$) and its vulnerability label $Y$ possess mutual information (i.e., the information conveyed by $Y$ provides insights into the value of $X$ (as well as $\tilde{X}$) and vice versa). Therefore, to guide the selection network in obtaining the important and label-relevant code statements $\tilde{X}$ (i.e., $\tilde{X}$ can predict the vulnerability $Y$ of $X$ correctly), inspired by \cite{learning-to-explain-l2x,van-ijcnn2021}, we apply to use the information theory principle \cite{TheoryofC, EofIT2006}. In particular, we maximize the mutual information between $\tilde{X}$ and $Y$ as mentioned in Eq. (\ref{eq:max_info}) with the intuition that by using the information from $Y$, the selection process $\zeta$ will be learned and enforced to obtain the most meaningful $\tilde{X}$ (i.e., $\tilde{X}$ can predict the vulnerability $Y$ of $X$ correctly).}

\vspace{-3mm}
{
\begin{equation}
\max_{\zeta}\,I(\tilde{X},Y).\label{eq:max_info}
\end{equation}
}
\vspace{-3mm}

Following \cite{EofIT2006}, we expand Eq. (\ref{eq:max_info}) further as the Kullback-Leibler divergence (i.e., it measures the difference in the information of two distributions) of the product of marginal distributions of $\tilde{X}$ and $Y$ from their joint distribution:

\vspace{-3mm}
{
\begin{align}
I(\tilde{X},Y)= \int p(\tilde{X},Y)\log\frac{p(\tilde{X},Y)}{p(\tilde{X})p(Y)}d\tilde{X}dY\label{eq:1sm}
\end{align}
}
\vspace{-3mm}

In practice, estimating mutual information is challenging as we typically only have access to samples but not the underlying distributions. Therefore, to estimate mutual information in Eq. (\ref{eq:1sm}), we employ a variational distribution $q(Y|\tilde{X})$ to approximate the posterior $p(Y|\tilde{X})$, hence deriving a variational lower bound of $I(\tilde{X}, Y)$, for which the equality holds if $q(Y|\tilde{X})=p(Y|\tilde{X})$, as follows:

\vspace{-2mm}
{
\begin{align}
I(\tilde{X},Y)\geq & \int p(Y,\tilde{X},X)\log\frac{q(Y|\tilde{X})}{p(Y)}dYd\tilde{X}dX\nonumber \\
= & E_{X,Y}E_{\tilde{X}|X}[\log q(Y|\tilde{X})]+\text{const}\label{eq:2sm}
\end{align}
}
\vspace{-2mm}

We model the variational distribution $q(Y|\tilde{X})$ presented in Eq. (\ref{eq:2sm}) by using a classifier implemented with a neural network $f(\tilde{X};\beta)$, which takes $\tilde{X}$ as input and outputs its corresponding label $Y$ (i.e., we view the classifier model as a distribution $q_{class}(Y|\tilde{X};\beta)$). 

\vspace{1mm}
The aim is then to learn both the selection model and the classifier by maximizing the following objective function:

\vspace{-3mm}
{
\begin{equation}
\underset{\alpha,\beta}{\text{max}}\left\{E_{X,Y}E_{\mathbf{z}\sim q_{sel}(\mathbf{z}|X;\alpha)}[\text{log}q_{class}(Y|X\odot\mathbf{z};\beta)]\right\}\label{eq:max_info_mutual_v2}
\end{equation}
}
\vspace{-3mm}

\paragraph{\textbf{Ensuring the selection of meaningful code statements}}

The joint training process between the classifier $f(.,\beta)$ and the selection network $g(.,\alpha)$ brings benefits for selecting the important and label-relevant features from the corresponding data. However, this process can also cause a potential limitation. \revise{In particular, the predictions of the classifier $f(.,\beta)$ may rely more on the features selected by the selection network $g(.,\alpha)$ than on the underlying information contained within those features. Consequently, the selected information (i.e., statements) may represent arbitrary subsets of the entire set of statements, rather than meaningful ones from the data.}

\vspace{-0mm}

To this end, we present an \textbf{\ourmechanism~} to ensure the learnable selection process $\zeta$ respecting the data distribution to select both the significant and label-relevant information of the data. In particular, in addition to learning the classifier jointly with the selection network as mentioned in Eq. (\ref{eq:max_info_mutual_v2}), inspired by \cite{realx2021}, we propose to learn the classifier model $f(.,\beta)$ disjointly to approximate the ground truth conditional distribution of $Y$ given subsets of $X$ denoted by $X_R$ where $X_{R}=X\odot \mathbf{r}$ with $\mathbf{r}\sim \text{MultiBernoulli}(0.5)$ (denoted by $\mathbf{r}\sim \text{B}(0.5)$ for short). \revise{This procedure aids in adjusting the classifier, enabling it to be influenced not only by the information obtained from the selection network $\zeta$ but also by the underlying information from the data when updating its parameter $\beta$.} This process is formalized as learning $q_{class}(.;\beta)$ to maximize:

\vspace{-3mm}
{
\begin{gather}
E_{X,Y}E_{\mathbf{r}\sim \text{B}(0.5)}\{\text{log}q_{class}(Y|X\odot\mathbf{r};\beta)\}\label{eq:realx}
\end{gather}
}
\vspace{-3mm}

\revise{To make this process, involving sampling operations from a Multi-Bernoulli distribution, remain continuous and differentiable during training, we employ the Gumbel-Softmax trick {\cite{jang2016categorical, MaddisonMT16}}. This technique relaxes each Bernoulli variable $r_{i} \in \mathbf{r}$ using the RelaxedBernoulli distribution function \cite{RelaxedBernoulli}.}

\vspace{1mm}
\subsubsection{\textbf{Innovative cluster-contrastive learning for improving source code data representation learning}}\label{sec:cluster_contrastive_learning}

\paragraph{\textbf{Motivation}}

Using contrastive learning \cite{Khosla2020} for improving data representations has been widely applied to computer vision domain applications from vision classification, detection, and segmentation \cite{NEURIPS2020_Kim, Wang_2021_ICCV, Sun_2021_CVPR, Du_2022_CVPR, Wang_2021_CVPR} to vision data out-of-distribution detection \cite{Sehwag21}. Its main idea is to encourage the data having the same label to share the close representations. 

\revise{In our study, we also aim to explore the potential of contrastive learning to enhance the learning process for source code data representation learning.} However, via our observations, it seems to be overdoing and not appropriate when straightforwardly applying contrastive learning to the source code data because even from the same vulnerability label, source code data can often consist of different hidden vulnerability patterns causing them vulnerable (in our experiment in Section \ref{sec:experiments}, we demonstrate this observation by comparing our \ourapp~method with one of the most effective OOD approaches using contrastive learning, namely SSD \cite{Sehwag21}). 

\begin{figure}[th]
\vspace{0mm}
\begin{centering}
\includegraphics[scale=0.5]{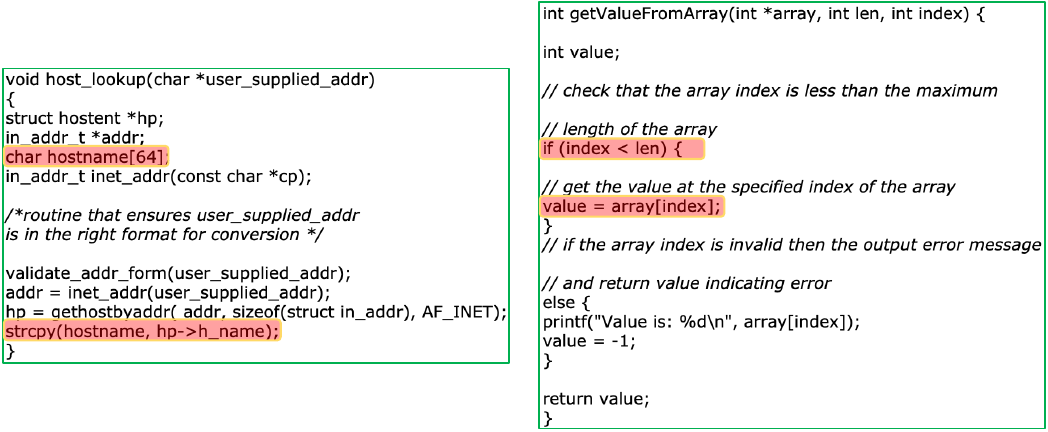}
\par\end{centering}
\vspace{0mm}
\caption{Two examples of vulnerability patterns causing the buffer overflow error \cite{cwe}. \textit{The left-hand function} shows an example of the buffer copy without checking the size of the input. \textit{The right-hand function} exhibits an example of the improper validation of an array index. It only verifies the array index against the maximum length, not the minimum value.} \label{fig:boe_update}
\vspace{0mm}
\end{figure}

As illustrated in Figure \ref{fig:boe_update}, despite sharing the same vulnerability $Y$ label (e.g., CWE categories, e.g., buffer overflow error), there are still several different hidden vulnerability patterns causing the source code data samples to be vulnerable. Consequently, although in the same vulnerability $Y$ label, the representations of vulnerable source code samples formed from different hidden vulnerability patterns need to be different. Otherwise, when they are from the same vulnerability patterns, their representations should be close or similar. \textit{To ensure these properties and leverage the power of contrastive learning to further improve the data representation learning process, we propose using innovative cluster-contrastive.}

\vspace{0mm}
\paragraph{\textbf{Innovative cluster-contrastive learning}\label{par:Clustered-spatial-contrastive}}

It is important to highlight that vulnerable source code data samples originating from the same vulnerability patterns tend to have similar $\tilde{X}$ and form the corresponding cluster in the latent space. Therefore, we first use clustering analysis (e.g., $k$-means) on the representations $\tilde{X}$ of the vulnerable source code samples  in the latent space to group vulnerable  samples  with the same vulnerability patterns into clusters before employing contrastive learning to force vulnerable data to become more similar in the same clusters and to be pushed far way in different clusters as follows:

\vspace{-5mm}
{\small{}
\begin{equation}
\mathcal{L}_{ccl} = \sum_{i\in I}1_{Y_{i}=1}\frac{-1}{\left|C(i)\right|}\sum_{c\in C(i)}\log\frac{\exp(\text{sim}({\tilde{X}_{i}},{\tilde{X}_{c}})/\tau)}{\sum\limits_{a\in A(i)}\exp(\text{sim}({\tilde{X}_{i}},{\tilde{X}_{a}})/\tau)} \label{eq:cluster_cl}
\end{equation}
}{\small\par}
\vspace{-1mm}

where $I\equiv\{1...m\}$ is a set of indices of input data in a specific mini-batch. $A(i)\equiv I\setminus\{i\}$ while $C(i)\equiv\{c\in A(i):\bar{Y}_{c}=\bar{Y}_{i}$ and $Y_{i}=1\}$ is the set of indices of vulnerable source code samples (labeled $1$) which are in the same cluster as $F_{i}$ except $i$, and $\left|C(i)\right|$ is its cardinality. Here, we apply $k$-means for each mini-batch and denote $\bar{Y}_{i}$ as the cluster label of the data sample $\tilde{X}_{i}$. For the $sim(.,.)$ function, we apply cosine similarity. Note that to form the corresponding vector for each $\tilde{X}_{i}$ in the latent space for calculating the cosine similarity, we simply concatenate all vectors where each vector stands for a representation of a code statement in $\tilde{X}_{i}$).

It is worth noting that cluster-contrastive learning described in Eq. (\ref{eq:cluster_cl}) helps enforce important properties of the source code data including (i) the vulnerable and non-vulnerable source code samples should have different representations, (ii) the vulnerable source code samples from different hidden vulnerability patterns are also encouraged to have different representations while (iii) the vulnerable source code samples in the similar hidden vulnerability patterns are trained to have close representations. Ensuring these properties helps improve the selection process. That helps boost the source code data representation learning facilitating out-of-distribution source code data identification.

\vspace{1mm}
\subsubsection{\textbf{A summary of our \ourapp~method}}

Algorithm \ref{alg:The-training-algorithm-prom} exhibits the details of our proposed \ourapp~method in the training and testing phases.

\RestyleAlgo{ruled}
\begin{algorithm*}[htbp]
\vspace{1mm}
\DontPrintSemicolon % Some LaTeX compilers require you to use \dontprintsemicolon instead
%\SetAlgoLined
%\SetAlgoVlined
\LinesNumbered

\KwIn{
An in-distribution source code dataset $D_{in}=\left\{ \left(X_{1},Y_{1}\right),\dots,(X_{N_{D_{in}}},Y_{N_{{D_{in}}}})\right\}$ (associated with one or multiple CWE categories) where each source code data sample $X_{i}$ consisting of $L$ code statements from $\mathbf{x}_{1}$ to $\mathbf{x}_{L}$ while its vulnerability $Y_{i}\in\left\{ 0,1\right\} $ (i.e., $1$: vulnerable and $0$: non-vulnerable) will be used during the training phase. An out-of-distribution CWE source code data category $D_{out}$ is used at the inference (testing) phase, denoted as $X_{test}$. \newline
We denote the number of training iterations $nt$; the mini-batch size $m$;
the trade-off hyper-parameter $\lambda$; and the number of clusters used in cluster-contrastive learning $K$. \newline
We randomly partition $D_{in}$ into the training set $D_{train}$ and the validation set $X_{val}$.
}
\BlankLine
We initialize the parameters $\alpha$ and $\beta$ of the selection model $\zeta$ parameterized by $g(.,\alpha)$ and the classifier model $f(.,\beta)$, respectively.
\BlankLine
\textbf{Training Phase}
\BlankLine
\For{$t=1$ to $nt$}
{
Choose a mini-batch of source code data samples denoted by $\{(X_{i},Y_{i})\}_{i=1}^{m}$ from $D_{train}$.
\BlankLine
Update the classifier parameter $\beta$ via minimizing the following cross-entropy loss $\mathcal{L}_{ce}$ $E_{X,Y}E_{\mathbf{r}\sim B(0.5)}[\mathcal{L}_{ce}(Y,f_{\beta}(X\odot\mathbf{r})]$ using the Adam optimizer \cite{KingmaB14}. Note that minimizing this function is equivalent to maximizing the objective function mentioned in Eq. (5).\;
\BlankLine
Update the classifier's parameter $\beta$ and the selection model parameter's $\alpha$ via minimizing the following objective function $E_{X,Y}E_{\mathbf{z}\sim q_{sel}(\mathbf{z}|X;\alpha)}[\mathcal{L}_{ce}(Y,f_{\beta}(X\odot\mathbf{z}))+\lambda\mathcal{L}_{ccl}]$ using the Adam optimizer. The $\mathcal{L}_{ccl}$ is described in Eq. (6).\;
\BlankLine
}

\revise{\textbf{Note that}, after the training phase, we aim to obtain the optimal selection network for the source code data representation learning. The trained selection network can select the most meaningful and label-relevant source code statements, forming the hidden vulnerability pattern to vulnerable data, for each source code data in the learned in-distribution $D_{in}$ to facilitate the model's ability to identify out-of-distribution CWE source code data.}

\BlankLine
\textbf{Testing Phase}
\BlankLine
We partition $D_{train}$ into $K$ clusters. In particular, we use $k$-means to cluster the entire training data representations in the latent space into $K$ clusters equal to the number of clusters used in the innovative cluster-contrastive learning (mentioned in Eq. (6)).
\BlankLine
$\forall X_{i}\in X_{test}$ (denoted as $X_{i\_test}$). 
We use Mahalanobis distance to calculate the outlier (out-of-distribution) score for each testing data sample $X_{i\_test}$ as follows
$s(X_{i\_test})=min_{k}(\tilde{X}_{i\_test}-\mu_{k})^{\mathrm{\top}}\Sigma_{k}^{-1}(\tilde{X}_{i\_test}-\mu_{k})$
where $\mu_{k}$ and $\Sigma_{k}$ are the sample mean and sample covariance of cluster $k^{th}\in K$.
In short, $s_{X_{test}}=score(\tilde{X}_{test},\mu_{X_{train}},\Sigma_{X_{train}})$
\BlankLine
$\forall X_{i}\in X_{val}$ (denoted as $X_{i\_val}$). 
We use Mahalanobis distance to calculate the outlier (out-of-distribution) score for each validation data sample $X_{i\_val}$ as follows
$s(X_{i\_val})=min_{k}(\tilde{X}_{i\_val}-\mu_{k})^{\mathrm{\top}}\Sigma_{k}^{-1}(\tilde{X}_{i\_val}-\mu_{k})$
where $\mu_{k}$ and $\Sigma_{k}$ are the sample mean and sample
covariance of cluster $k$-th $\in K$.
In short, $s_{X_{val}}=score(G_{X_{val}},\mu_{X_{train}},\Sigma_{X_{train}})$
\BlankLine
Calculate $s_{X_{val}}$ (threshold) at $TPR=95\%$. (Note that the threshold value is chosen at $TPR=95\%$ as commonly used in OOD detection methods)
\BlankLine
\eIf{($s(X_{i\_test})>s_{X_{val}}$)}
{
	$X_{i\_test}$ is an out-of-distribution data.
}
{
	$X_{i\_test}$ is an in-distribution data.
}
\BlankLine
\KwOut{$\forall X_{i}\in X_{test}$ (denoted as $X_{i\_test}$) whether $X_{i\_test}$ is out-of-distribution data sample.}\caption{The algorithm of our proposed \ourapp~method for out-of-distribution source code data identification.\label{alg:The-training-algorithm-prom}}
\vspace{0mm}
\end{algorithm*}

\paragraph{\textbf{The training phase}}
In brief, during the training process, we update the parameters $\alpha$ and $\beta$ of the selection model $\zeta$ parameterized by $g(.,\alpha)$ and the classifier model $f(.,\beta)$ simultaneously through the objective functions presented in Eqs. (\ref{eq:max_info_mutual_v2}, \ref{eq:realx}, and \ref{eq:cluster_cl}). \revise{After the training phase, our objective is to leverage the optimal trained selection network to create distinct representations in the latent space for source code data stemming from different hidden vulnerability patterns. This ensures they are positioned far apart, while source code data sharing similar hidden vulnerability patterns remain closely clustered. This elegant process effectively facilitates the identification of out-of-distribution source code data.}

\vspace{1mm}
\paragraph{\textbf{The inference phase}}
Inspired by \cite{Lee18b, Winkens2020, Sehwag21}, we use the cluster-conditioned OOD detection method (i.e., applying the Mahalanobis distance to calculate the outlier score $s(X)$ for each testing sample $X$) in the latent space for OOD source code data identification.
The procedure for a cluster-conditioned detection method is as follows: We first partition the features of in-distribution training data into $K$ clusters equal to the number of clusters used in cluster-contrastive learning. We then compute the outlier score $s(X)$ of each testing input $X$ as $s(X)=min_{k}(\tilde{X}-\mu_{k})^{\mathrm{\top}}\Sigma_{k}^{-1}(\tilde{X}-\mu_{k})$ where $\mu_{k}$ and $\Sigma_{k}$ are the sample mean and sample covariance of each cluster $k$ in $K$ clusters. We finally compare this outlier score with a threshold value to decide if the corresponding input sample is OOD.

\vspace{3mm}
\section{Experiments}\label{sec:experiments}

\subsection{\textbf{Experimental designs}}

The key goal of our experiments is to evaluate our \ourapp~method and compare it with state-of-the-art baselines for out-of-distribution source code data identification. Below, we present the research questions of our paper.

\vspace{1mm}
\textbf{(RQ1) \rqone}

For AI-based systems, being able to detect data that is out-of-distribution can be critical to maintaining safe and reliable predictions. Although achieving promising performances, current state-of-the-art OOD methods are mostly applied to computer vision and medical diagnosis applications. Automated deep learning-based approaches for OOD identification to the source code data have not yet been well studied and explored. 

To this end, we propose one of the first methods for OOD source code data identification. We demonstrate that by taking into account source code characteristics and leveraging the correlations of the potentially hidden vulnerability patterns inside and between the source code data samples from the learned in-distribution $D_{in}$, we can significantly improve the source code data representation learning for boosting OOD source code data identification.

\vspace{1mm}
\textbf{(RQ2) \rqtwo}

\revise{In practice, the in-distribution $D_{in}$ may encompass source code data from one or multiple CWE categories. Therefore, in this research question, we investigate the performance of our \ourapp~method in various cases regarding different numbers of CWE categories (i.e., without loss of generality, we assume there are two and three different CWE categories) presenting in the in-distribution source code data  $D_{in}$ and observe how the model performs compared to the baselines.}

\vspace{1mm}
\textbf{(RQ3) \rqthree}

\revise{We investigate the effectiveness of the cluster-contrastive learning (considering source code characteristics and semantic relationships inside and between code data) and \ourmechanism~(ensuring the selection of significant source code statements) in enhancing the entire source code data representation learning process, helping boost the model's ability for OOD source code data identification.}

\vspace{1mm}
\subsubsection{\textbf{Studied datasets}}\label{sec:studied_data_sets}

We conducted experiments on many different real-world (CWE) types extracted from \textit{an up-to-date and big C/C++ dataset, namely DiverseVul} provided by \cite{chen2023diversevul} which contains many types of CWE categories. The DiverseVul dataset contains 18,945 vulnerable functions and 330,492 non-vulnerable functions covering 150 CWEs.

In the scope of our paper, we focus on conducting our experiments on the top most dangerous and popular common weakness enumeration (CWE) categories \cite{chen2023diversevul} including CWE-119 (Improper Restriction of Operations within the Bounds of a Memory Buffer), CWE-20 (Improper Input Validation), CWE-125 (Out-of-bounds Read), CWE-200 (Exposure of Sensitive Information to an Unauthorized Actor), CWE-787 (Out-of-bounds Write), CWE-416 (Use After Free), CWE-269 (Improper Privilege Management), CWE-94 (Improper Control of Generation of Code (Code Injection)), CWE-190 (Integer Overflow or Wraparound), CWE-264 (Permissions, Privileges, and Access Controls), CWE-863 (Incorrect Authorization), CWE-862 (Missing Authorization), and CWE-287 (Improper Authentication).

\subsubsection{\textbf{Data processing and embedding}}

We preprocessed the datasets before injecting them into our proposed \ourapp~method and baselines. In particular, we standardized the source code data samples by (i) removing comments, blank lines, and non-ASCII characters, and (ii) mapping user-defined variables to symbolic variable names (e.g., \emph{var1} and \emph{var2}) and user-defined
functions to symbolic function names (e.g., \emph{func1} and \emph{func2}) using Treesitter \cite{Treesitter}.  We also replaced strings with a generic \emph{"str"} token.

We then embedded source code statements into vectors. For instance, consider the following statement (written in C/C++ programming language) \textit{for(var1=0;var1$<$10;var1++}, we tokenize this statement into a sequence of tokens (i.e., \textit{for,(,var1,=,0,;,var1,$<$,10,;,var1,++,)}), and then we used a 150-dimensional Embedding layer followed by a Dropout layer (with a dropped fixed probability $p=0.2$), a 1D convolutional layer (with the filter size $150$ and kernel size $3$), and a 1D max pooling layer to encode each source code statement in a source code function $X$. Note that we utilize the commonly used values for these hyperparameters. Finally, a mini-batch of functions in which each function consisting of $L$ encoded statements was fed to the deep learning-based models. It is worth noting that the Embedding and 1D convolutional layers are learnable during the training process.

It is worth noting that from the used datasets, the length ($L$) of each function is padded or truncated to $100$ source code statements  (i.e., we base on the quantile values of the source code data’ length to decide the length of each source code data sample). Specifically, more than 95\% of the functions consist of 100 or fewer source code statements. Furthermore, we observe that almost all important information relevant to the vulnerability of each source code data sample (e.g., a function) lies in the first $100$ source code statements.

\vspace{1mm}
\subsubsection{\textbf{Baseline methods}}

\textbf{The main baselines} of our proposed \ourapp~method are some effective and state-of-the-art methods for out-of-distribution detection including Standard DNN \cite{Hendrycks2017dnn}, Outlier Exposure \cite{Hendrycks18}, and SSD \cite{Sehwag21}. These methods were originally applied to computer vision data applications. To make these methods able to be applied to the source code data, we keep the principles and make some modifications on the network for data representation learning as used in our method. There are other methods, e.g., \cite{Hendrycks2019, Winkens2020, Tack2020}, recently introduced for out-of-distribution detection. However, these methods were operated based on the nature of the vision data, so they are not applicable to the source code data.

\vspace{1mm}
\textbf{The additional baselines} of our \ourapp~method are some state-of-the-art SVD methods including VulDeePecker \cite{VulDeePecker2018}, CodeBERT \cite{CodeBERT2020}, and ReGVD \cite{ReGVD2022}. We investigate how these effective SVD methods perform for identifying OOD source code data. The rationale is that an effective model should produce high-quality data representations that also facilitate distinguishing the data between in-distribution and OOD.

\vspace{1mm}
We briefly summarize the baselines as follows:
\begin{itemize}
    \item \textbf{Standard DNN} \cite{Hendrycks2017dnn}. It is a baseline utilizing probabilities from softmax distributions for detecting misclassified and out-of-distribution examples in neural networks.\vspace{1mm}
    \item \textbf{Outlier Exposure} \cite{Hendrycks18}. It enables OOD (anomaly) detectors to generalize and detect unseen anomalies by training anomaly detectors against an auxiliary dataset of outliers disjoint from the testing dataset (i.e., OOD data).\vspace{1mm}
    \item \textbf{SSD} \cite{Sehwag21}. An outlier detector method utilizes contrastive representation learning followed by a Mahalanobis distance-based detection in the feature space for outlier (out-of-distribution) detection.\vspace{1mm}
    \item \textbf{VulDeePecker} \cite{VulDeePecker2018}. It is one of the most effective deep learning-based software vulnerability detection methods leveraging the power of bi-directional long-short-term memory (LSTM) networks \cite{hochreiter1997long-short}.\vspace{1mm}
    \item \textbf{CodeBERT} \cite{CodeBERT2020}. A pre-trained model (based on the Transformer's encoder framework \cite{transformers-Vaswani17}) specializes in the programming language. Via CodeBERT, we aim to investigate how large language models work in the context of out-of-distribution source code data identification.\vspace{1mm}
    \item \textbf{ReGVD} \cite{ReGVD2022}. It is an effective Graph neural network-based model (i.e., using Graph convolutional networks \cite{GCNN2017} and Gated graph neural networks \cite{Gated2016}) for the software vulnerability detection problem.
\end{itemize}

\vspace{1mm}
\subsubsection{\textbf{Model's configurations}}

For the main baselines of our \ourapp~method including Standard DNN \cite{Hendrycks2017dnn}, Outlier Exposure \cite{Hendrycks18}, and SSD \cite{Sehwag21}. These methods are popular and state-of-the-art approaches for out-of-distribution detection applied in the computer vision domain. To make them applicable for out-of-distribution source code vulnerability detection, we keep the principle of these methods and use the same data embedding process for handling the sequential source code data as used in our method. For the additional baseline approaches, popular and state-of-the-art SVD methods (i.e., VulDeePecker \cite{VulDeePecker2018}, CodeBERT \cite{CodeBERT2020}, and ReGVD \cite{ReGVD2022}), we use the architecture proposed in the corresponding papers.

To our \ourapp~method, for the $g\left(\cdot;\alpha\right)$
and $f\left(\cdot;\beta\right)$ networks, we used deep feed-forward neural networks having three and two hidden layers with the size of each hidden layer in $\left\{100,300\right\} $. The dense hidden layers are followed by a ReLU function as nonlinearity and Dropout \cite{srivastava14a} with a retained fixed probability $p=0.8$ as regularization. The last dense layer of the $g\left(\cdot;\alpha\right)$ network for learning a discrete distribution is followed by a sigmoid function while the last dense layer of the $f\left(\cdot;\beta\right)$ network is followed by a softmax function for predicting. The number of chosen clusters guiding the computation of the innovative cluster-contrastive learning mentioned is set in $\{1, 3, 5, 7, 9\}$. The trade-off hyper-parameter $\lambda$ representing the weight of the innovative cluster-contrastive learning is in $\{10^{-2},10^{-1},10^{0}\}$ while the scalar temperature $\tau$ is in $\{0.5, 1.0\}$. The temperature $\nu$ for the Gumbel softmax distribution is also set in $\{0.5, 1.0\}$. Note that we utilize the commonly used values for these hyperparameters.

For our \ourapp~method and baselines, we employed
the Adam optimizer \cite{KingmaB14} with an initial learning rate equal to $10^{-3}$, while the mini-batch size is set to $128$. For the training process, we split the data of each in-distribution data into two random partitions. The first partition contains 80\% for training, the second partition contains 20\% for validation. For each in-distribution data, we used $10$ epochs for the training process. We additionally applied gradient clipping regularization to prevent over-fitting. For each method, we ran the corresponding model several times and reported the averaged FPR (at TPR 95\%), AUROC, and AUPR measures. We ran our experiments in Python using Tensorflow \cite{abadi2016tensorflow} for the used methods on a 13th Gen Intel(R) Core(TM) i9-13900KF having 24 CPU Cores at 3.00 GHz with 32GB RAM, integrated Gigabyte RTX 4090 Gaming OC 24GB. Some baseline methods (i.e., CodeBERT \cite{CodeBERT2020} and ReGVD \cite{ReGVD2022}) were written using Pytorch \cite{pytorch2019}. For these baselines, we followed the source code samples published by the authors.

\vspace{1mm}
\subsubsection{\textbf{Measures}}

To measure the performance of our \ourapp~method and baselines, we use three main metrics, i.e., FPR (at TPR 95\%), AUROC, and AUPR, commonly used in out-of-distribution (OOD) detection methods, as noted in \cite{Hendrycks18, Ren19, Sehwag21}. In particular, the FPR (at TPR 95\%) metric calculates how many percent of OOD data are predicted as in-distribution data. \textit{For the FPR measure, the smaller value is better}. The AUROC (the area under the receiver operating characteristic) metric presents the model’s ability to discriminate between the in-distribution and OOD data while the AUPR (the area under the precision-recall curve) metric measures if the model can find all OOD data without accidentally marking any in-distribution data as OOD data. \textit{For the AUROC and AUPR measures, the higher value is better}.

\vspace{1mm}
\subsection{\textbf{Experimental results}}\label{sec:exp_results}

\textbf{RQ1: \rqone}
\vspace{0mm}
\paragraph{\textbf{Approach}}

We compare the performance of our \ourapp~ method with the baselines including Standard DNN \cite{Hendrycks2017dnn}, Outlier Exposure \cite{Hendrycks18}, SSD \cite{Sehwag21}, VulDeePecker \cite{VulDeePecker2018}, CodeBERT \cite{CodeBERT2020}, and ReGVD \cite{ReGVD2022} in the task of out-of-distribution source code data identification using FPR (at TPR 95\%), AUROC, and AUPR measures.

As mentioned in Section \ref{sec:studied_data_sets}, we conduct our experiments on the top most dangerous and popular software common weakness enumeration (CWE). In the training phase, we alternatively use source code data from each CWE category as in-distribution data to train our \ourapp~method and the baselines. In the inference (testing) phase, for each trained model on a specific in-distribution CWE source code category, we use the remaining source code data from other CWE categories as out-of-distribution data to evaluate the model's performance. For example, we train each model using source code data from the CWE-269 category as in-distribution data. In the testing phase, we use source code data from other CWE categories (e.g., CWE-20 and CWE-200) as out-of-distribution data to evaluate the model performance. We then do the same process for source code data from other CWE categories. 

\textit{It is worth noting that following the setting of the out-of-distribution identification problem, we do not use any information from out-of-distribution CWE source code data categories in the training phase.} We only use out-of-distribution CWE categories in the inference (testing) process for evaluating the models' performance. \revise{In practice, vulnerable data falling under out-of-distribution CWE categories pose a significantly higher risk to security systems compared to non-vulnerable data. Therefore, in our experiments, we primarily investigate the performance of our \ourapp~method and the baselines on vulnerable data belonging to OOD CWE categories.}

\begin{table}[ht]
\vspace{1.5mm}
\caption{\revise{The results of our \ourapp~method and baselines for the FPR (at TPR 95\%), AUROC, and AUPR measures on the vulnerable source code of each OOD CWE category corresponding with specific in-distribution (ID) data.  (The best results are in \textbf{bold} while the second highest results are in \underline{underline}. The numbers highlighted in blue represent the improvements of our method over the second-best baseline.)}}\label{tab:my_label1t}
\centering{}
\vspace{0mm}
\resizebox{0.95\columnwidth}{!}{
%{|c|c|r|r|r|}
\begin{tabular}{ccrrr}
\hline 
%\textbf{ID and OOD} & \textbf{Methods} & \textbf{FPR} $\downarrow$ & \textbf{AUROC} $\uparrow$ & \textbf{AUPR} $\uparrow$
\textbf{ID and OOD} & \textbf{Methods} & \textbf{FPR} $\downarrow$ & \textbf{AUROC} $\uparrow$ & \textbf{AUPR} $\uparrow$
\tabularnewline
\hline 
%\hline		

\multirow{7}{*}{CWE287 vs. CWE416}
 &  Standard DNN & 86.30\%	& 67.93\% & 84.60\% 
 \tabularnewline
%\cline{2-5} \cline{3-5} \cline{4-5} \cline{5-5} 
 &  Outlier Exposure &  79.30\%	& \underline{68.93\%} & \underline{86.13\%}
 \tabularnewline
%\cline{2-5} \cline{3-5} \cline{4-5} \cline{5-5} 
 &  SSD & 84.78\%	& 66.71\% & 84.67\%
 \tabularnewline
%\cline{2-5} \cline{3-5} \cline{4-5} \cline{5-5} 
 &  VulDeePecker & \underline{77.63\%}	& 68.20\% & 85.61\%
 \tabularnewline
%\cline{2-5} \cline{3-5} \cline{4-5} \cline{5-5} 
 &  CodeBERT &  88.57\% & 68.25\% & 85.49\%
 \tabularnewline
%\cline{2-5} \cline{3-5} \cline{4-5} \cline{5-5} 
 &  ReGVD &  90.40\%	& 58.58\% & 81.88\%
 \tabularnewline
\cline{2-5} \cline{3-5} \cline{4-5} \cline{5-5} 

   & \multirow{2}{*}{\ourapp~(Ours)} & \textbf{69.71\%} & \textbf{76.75\%} & \textbf{90.57\%}\tabularnewline
%\cline{3-5} \cline{4-5} \cline{5-5} 
 &  & \textcolor{blue}{($\downarrow$ 7.92\%)} & \textcolor{blue}{($\uparrow$ 7.82\%)} & \textcolor{blue}{($\uparrow$ 4.44\%)}\tabularnewline
 
\hline 
%\hline 	
 
 \multirow{7}{*}{CWE269 vs. CWE200}
 &  Standard DNN & \underline{79.36\%}	& 69.28\%	& 78.75\%
 \tabularnewline
%\cline{2-5} \cline{3-5} \cline{4-5} \cline{5-5} 
 &  Outlier Exposure &  80.30\%	& 70.05\%	& 78.49\%
 \tabularnewline
%\cline{2-5} \cline{3-5} \cline{4-5} \cline{5-5} 
 &  SSD & 79.92\%	& 72.70\%	& 78.86\%
 \tabularnewline
%\cline{2-5} \cline{3-5} \cline{4-5} \cline{5-5} 
 &  VulDeePecker & 83.90\%	& \underline{74.73\%}	& \underline{81.62\%}
 \tabularnewline
%\cline{2-5} \cline{3-5} \cline{4-5} \cline{5-5} 
 &  CodeBERT & 88.43\%	& 65.64\%	& 74.55\%
 \tabularnewline
%\cline{2-5} \cline{3-5} \cline{4-5} \cline{5-5} 
 &  ReGVD  & 93.74\%	& 61.01\%	& 69.55\%
 \tabularnewline
\cline{2-5} \cline{3-5} \cline{4-5} \cline{5-5} 

 & \multirow{2}{*}{\ourapp~(Ours)} & \textbf{76.89\%} & \textbf{79.33\%} & \textbf{85.32\%}\tabularnewline
%\cline{3-5} \cline{4-5} \cline{5-5} 
 &  & \textcolor{blue}{($\downarrow$ 2.47\%)} & \textcolor{blue}{($\uparrow$ 4.60\%)} & \textcolor{blue}{($\uparrow$ 3.70\%)}\tabularnewline

 \hline
 %\hline

\multirow{7}{*}{CWE863 vs. CWE287}
&  Standard DNN & 50.00\%	& 80.28\%	& 78.26\%
\tabularnewline
%\cline{2-5} \cline{3-5} \cline{4-5} \cline{5-5} 
&  Outlier Exposure & 71.21\% &  79.29\% &  76.57\%  
\tabularnewline
%\cline{2-5} \cline{3-5} \cline{4-5} \cline{5-5} 
&  SSD & \underline{46.97\%}	& \underline{82.55\%}	& \underline{83.78\%}
\tabularnewline
%\cline{2-5} \cline{3-5} \cline{4-5} \cline{5-5} 
&  VulDeePecker & 48.48\%	& 79.20\%	& 76.53\%
\tabularnewline
%\cline{2-5} \cline{3-5} \cline{4-5} \cline{5-5} 
&  CodeBERT  & 53.85\%	& 78.54\%	& 78.43\%
\tabularnewline
%\cline{2-5} \cline{3-5} \cline{4-5} \cline{5-5} 
&  ReGVD & 66.15\%	& 80.42\%	& 77.84\%
\tabularnewline
\cline{2-5} \cline{3-5} \cline{4-5} \cline{5-5} 

&  \multirow{2}{*}{\ourapp~(Ours)}  & \textbf{33.33\%} & \textbf{87.06\%} & \textbf{85.99\%}
\tabularnewline
%\cline{3-5} \cline{4-5} \cline{5-5} 
 &  & \textcolor{blue}{($\downarrow$ 13.64\%)} & \textcolor{blue}{($\uparrow$ 4.51\%)} & \textcolor{blue}{($\uparrow$ 2.21\%)}\tabularnewline
\hline

 %\hline 	
 
 \multirow{7}{*}{CWE94 vs. CWE20}
 &  Standard DNN & \underline{73.26\%}	& 75.60\%	& \underline{95.60\%}
 \tabularnewline
%\cline{2-5} \cline{3-5} \cline{4-5} \cline{5-5} 
 &  Outlier Exposure &  74.44\%	& \underline{75.83\%}	& 95.28\%
 \tabularnewline
%\cline{2-5} \cline{3-5} \cline{4-5} \cline{5-5} 
 &  SSD & 85.70\%	& 70.40\%	& 94.17\%
 \tabularnewline
%\cline{2-5} \cline{3-5} \cline{4-5} \cline{5-5} 
 &  VulDeePecker & 77.85\%	& 73.79\%	& 95.54\%
 \tabularnewline
%\cline{2-5} \cline{3-5} \cline{4-5} \cline{5-5} 
 &  CodeBERT & 87.55\%	& 74.93\%	& 94.58\%
 \tabularnewline
%\cline{2-5} \cline{3-5} \cline{4-5} \cline{5-5} 
 &  ReGVD  & 94.74\%	& 66.98\%	& 92.55\%
 \tabularnewline
\cline{2-5} \cline{3-5} \cline{4-5} \cline{5-5} 

  & \multirow{2}{*}{\ourapp~(Ours)} &  \textbf{51.70\%}	& \textbf{84.37\%} & \textbf{97.72\%}\tabularnewline
%\cline{3-5} \cline{4-5} \cline{5-5} 
 &  & \textcolor{blue}{($\downarrow$ 21.56\%)} & \textcolor{blue}{($\uparrow$ 8.54\%)} & \textcolor{blue}{($\uparrow$ 2.12\%)}\tabularnewline

\hline 	
 
\multirow{7}{*}{CWE190 vs. CWE119}
&  Standard DNN & 85.54\%	& 69.97\%	& 60.47\%
\tabularnewline
%\cline{2-5} \cline{3-5} \cline{4-5} \cline{5-5} 
&  Outlier Exposure &  82.69\%	& 69.93\% & 60.22\%
\tabularnewline
%\cline{2-5} \cline{3-5} \cline{4-5} \cline{5-5} 
&  SSD & \underline{77.87\%}	& \underline{74.46\%}	& \underline{66.04\%}
\tabularnewline
%\cline{2-5} \cline{3-5} \cline{4-5} \cline{5-5} 
&  VulDeePecker & 85.17\%	& 70.01\%	& 59.79\%
\tabularnewline
%\cline{2-5} \cline{3-5} \cline{4-5} \cline{5-5} 
&  CodeBERT  & 90.72\%	& 68.74\%	& 55.74\%
\tabularnewline
%\cline{2-5} \cline{3-5} \cline{4-5} \cline{5-5} 
&  ReGVD & 96.66\%	& 50.50\%	& 41.33\%
\tabularnewline
\cline{2-5} \cline{3-5} \cline{4-5} \cline{5-5} 

&  \multirow{2}{*}{\ourapp~(Ours)}  & \textbf{68.85\%} & \textbf{77.59\%} & \textbf{72.42\%}
\tabularnewline
%\cline{3-5} \cline{4-5} \cline{5-5} 
 &  & \textcolor{blue}{($\downarrow$ 9.02\%)} & \textcolor{blue}{($\uparrow$ 3.13\%)} & \textcolor{blue}{($\uparrow$ 6.38\%)}\tabularnewline
\hline	

\multirow{7}{*}{CWE269 vs. CWE20}
 &  Standard DNN & \underline{72.96\%}	& 71.90\%	& 91.43\%
 \tabularnewline
%\cline{2-5} \cline{3-5} \cline{4-5} \cline{5-5} 
 &  Outlier Exposure &  77.33\%	& \underline{73.29\%}	& \underline{91.62\%}
 \tabularnewline
%\cline{2-5} \cline{3-5} \cline{4-5} \cline{5-5} 
 &  SSD & 85.33\%	& 69.69\%	& 89.44\%
 \tabularnewline
%\cline{2-5} \cline{3-5} \cline{4-5} \cline{5-5} 
 &  VulDeePecker & 79.63\%	& 71.23\%	& 90.33\%
 \tabularnewline
%\cline{2-5} \cline{3-5} \cline{4-5} \cline{5-5} 
 &  CodeBERT & 91.40\%	& 67.44\%	& 87.62\%
 \tabularnewline
%\cline{2-5} \cline{3-5} \cline{4-5} \cline{5-5} 
 &  ReGVD & 92.51\%	& 62.60\%	& 86.31\%
 \tabularnewline
\cline{2-5} \cline{3-5} \cline{4-5} \cline{5-5} 

 & \multirow{2}{*}{\ourapp~(Ours)} & \textbf{64.30\%} & \textbf{79.51\%} & \textbf{93.95\%}\tabularnewline
%\cline{3-5} \cline{4-5} \cline{5-5} 
 &  & \textcolor{blue}{($\downarrow$ 8.66\%)} & \textcolor{blue}{($\uparrow$ 6.22\%)} & \textcolor{blue}{($\uparrow$ 2.33\%)}\tabularnewline
 
 \hline
%\hline
 
\hline
\end{tabular}}
\vspace{-3mm}
\end{table}

\vspace{1mm}
\paragraph{\textbf{Quantitative results}}

The experimental results in Table 1 demonstrate the superiority of our \ourapp~method over the baseline approaches. In particular, our method obtains much higher performances in all of the used measures including FPR, AUROC, and AUPR with a high margin for almost all cases of the in-distribution and out-of-distribution CWE source code data. In general, on average, our method achieves significantly higher performances than the baselines from around 14.20\%, 6.67\%, and 5.12\% on the FPR, AUROC, and AUPR measures, respectively, across the used in-distribution and out-of-distribution CWE source code data categories.

\vspace{1mm}
\paragraph{\textbf{Qualitative results}}
To further demonstrate the advancement and effectiveness of our \ourapp~method for OOD source code data identification compared to the baselines, we visualize the representations of both in-distribution and out-of-distribution source code data in the feature space. The visualizations help us investigate whether our \ourapp~method can successfully learn the characteristics and leverage the semantic relationships inside and between source code data \revise{to effectively distinguish between in-distribution and out-of-distribution source code data samples, facilitating addressing the OOD source code data identification problem.}

\begin{figure*}[ht]
\vspace{0mm}
\begin{centering}
\includegraphics[width=0.95\textwidth]{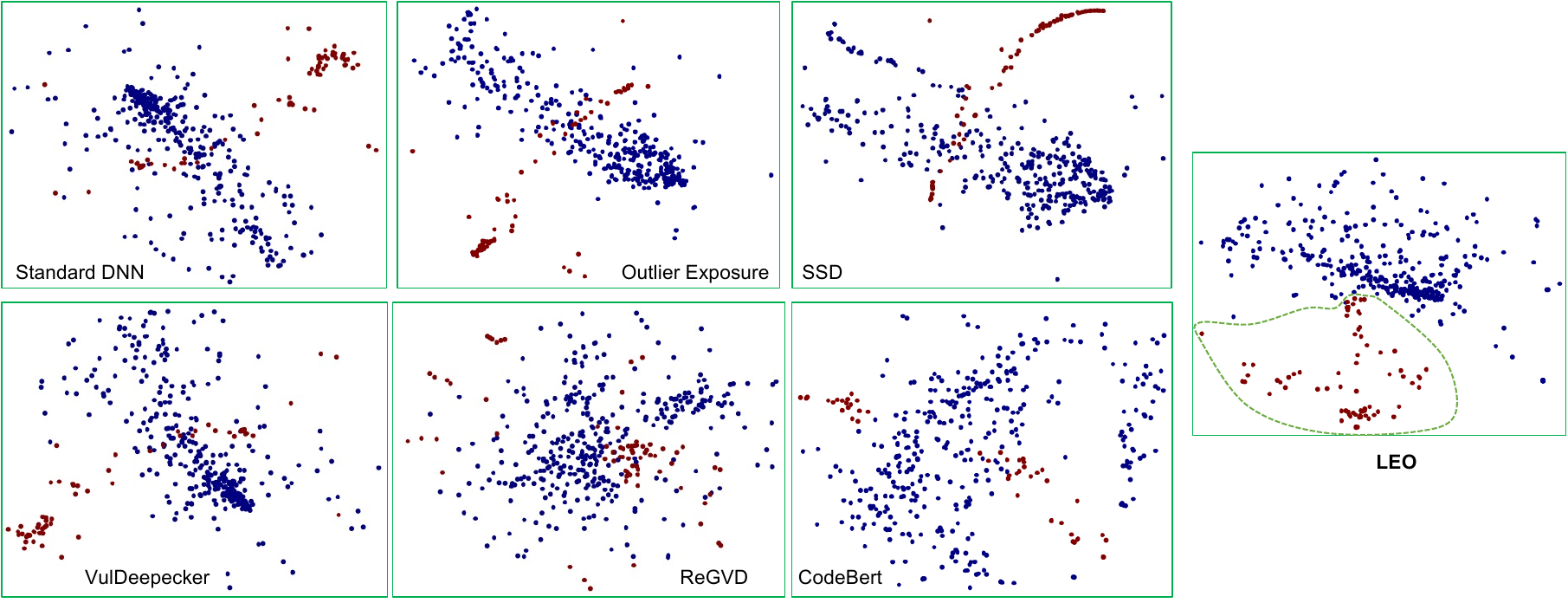}
\par\end{centering}\vspace{-1mm}
\caption{A 2D t-SNE projection for the data representation distribution of the in-distribution data (blue color) and the out-of-distribution vulnerable data (red color) in the latent space (i.e., where source code data from CWE863 and CWE287 categories are used as in-distribution data and out-of-distribution data, respectively) of our proposed \ourapp~method and the baselines.} \label{fig:t-SNE}
\vspace{0mm}
\end{figure*}

We use a t-SNE \cite{tSNE2008} projection, with perplexity equal to $300$, to visualize the distributions of the in-distribution and out-of-distribution data in the feature space. As illustrated in Figure \ref{fig:t-SNE}, it becomes apparent that our \ourapp~method significantly enhances the separation of the distributions between in-distribution and out-of-distribution source code data in the latent space, compared to the baselines. These visual representations provide compelling evidence of the robustness of our \ourapp~method in boosting the learning of source code data representations, thus facilitating the identification of out-of-distribution CWE source code data.

\vspace{3mm}
\colorbox{gray!20}{
\begin{minipage}{0.91\columnwidth}
\textbf{In conclusion for RQ1}: The quantitative results in Table \ref{tab:my_label1t} on three main measures (i.e., FPR (at TPR 95\%), AUROC, and AUPR) and the qualitative visualizations depicted in Figure \ref{fig:t-SNE} show the effectiveness and advancement of our \ourapp~method in achieving significantly higher performances for identifying OOD CWE source code data categories over the baselines. In short, our method achieves higher performances than the baselines from around 13.77\%, 7.88\%, and 4.84\% on average to the FPR, AUROC, and AUPR measures, respectively, on the used in-distribution and OOD CWE categories.
\end{minipage}
}
\vspace{3mm}

\vspace{0mm}

\begin{figure*}[ht]
\vspace{0mm}
\begin{centering}
\includegraphics[width=0.93\textwidth]{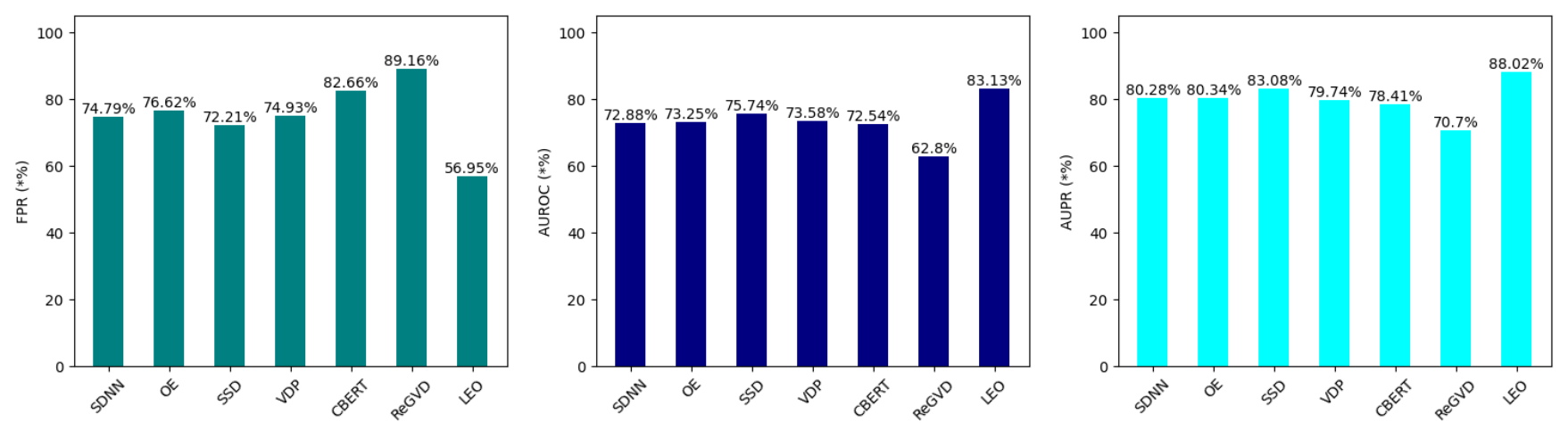}
\par\end{centering}\vspace{1mm}
\caption{The average results for the FPR (at TPR 95\%), AUROC, and AUPR measures, respectively, of our \ourapp~method and baselines in all cases of the in-distribution and out-of-distribution CWE categories mentioned in Tables \ref{tab:my_label1t} and \ref{tab:my_label23t}. Note that for the FPR measure, the smaller value is better while for the AUROC, and AUPR measures, the higher value is better. We denote the Standard DNN, Outlier Exposure, VulDeePecker, and CodeBERT methods as SDNN, OE, VDP, and CBERT for short.}\label{fig:averageresults}
\vspace{-3mm}
\end{figure*}%

\vspace{1mm}
\textbf{RQ2: \rqtwo}

\paragraph{\textbf{Approach}}

This research question reflects different reality cases where the in-distribution source code data are from more than one (as presented in Table \ref{tab:my_label1t}) CWE category. Therefore, we further investigate our proposed \ourapp~method's performance in cases of different CWE categories existing in the in-distribution source code data and observe how the model performs compared to the baselines.

\begin{table}[ht]
\vspace{1.5mm}
\caption{\revise{The results of our \ourapp~method and baselines for the FPR (at TPR 95\%), AUROC, and AUPR measures on the vulnerable source code of each OOD CWE category corresponding with specific in-distribution (ID) data. (The best results are in \textbf{bold} while the second highest results are in \underline{underline}. The numbers highlighted in blue represent the improvements of our method over the second-best baseline.)}}\label{tab:my_label23t}
\centering{}
\vspace{-1mm}
\resizebox{0.95\columnwidth}{!}{
% {|c|c|r|r|r|}
\begin{tabular}{ccrrr}
\hline 
%\textbf{ID and OOD} & \textbf{Methods} & \textbf{FPR} $\downarrow$ & \textbf{AUROC} $\uparrow$ & \textbf{AUPR} $\uparrow$
\textbf{ID and OOD} & \textbf{Methods} & \textbf{FPR} $\downarrow$ & \textbf{AUROC} $\uparrow$ & \textbf{AUPR} $\uparrow$
\tabularnewline
\hline 
 
\multirow{7}{*}
 &  Standard DNN & 54.55\%	& 80.00\%	& 64.71\%
 \tabularnewline
%\cline{2-5} \cline{3-5} \cline{4-5} \cline{5-5} 
 &  Outlier Exposure &  57.58\%	& 80.68\%	& 65.78\%
 \tabularnewline
%\cline{2-5} \cline{3-5} \cline{4-5} \cline{5-5} 
{CWE863+862}
 &  SSD & \underline{48.48\%}	& 89.22\%	& \underline{71.28\%}
 \tabularnewline
%\cline{2-5} \cline{3-5} \cline{4-5} \cline{5-5} 
{vs. CWE287}
 &  VulDeePecker & 60.61\%	& 82.48\%	& 56.61\%
 \tabularnewline
%\cline{2-5} \cline{3-5} \cline{4-5} \cline{5-5} 
 &  CodeBERT & 63.08\%	& \underline{89.60\%}	& 64.08\%
 \tabularnewline
%\cline{2-5} \cline{3-5} \cline{4-5} \cline{5-5} 
 &  ReGVD  & 83.33\%	& 67.89\%	& 39.98\%
 \tabularnewline
\cline{2-5} \cline{3-5} \cline{4-5} \cline{5-5} 

 & \multirow{2}{*}{\ourapp~(Ours)} &  \textbf{37.88\%}	& \textbf{90.48\%} & \textbf{78.69\%}\tabularnewline
%\cline{3-5} \cline{4-5} \cline{5-5} 
 &  & \textcolor{blue}{($\downarrow$ 10.60\%)} & \textcolor{blue}{($\uparrow$ 0.88\%)} & \textcolor{blue}{($\uparrow$ 7.41\%)}\tabularnewline
 
\hline 
%\hline 
 
\multirow{7}{*}
 &  Standard DNN & 81.48\%	& 73.10\% & 86.88\% 
 \tabularnewline
%\cline{2-5} \cline{3-5} \cline{4-5} \cline{5-5} 
 &  Outlier Exposure &  75.41\%	& 75.52\% & 88.55\%
 \tabularnewline
%\cline{2-5} \cline{3-5} \cline{4-5} \cline{5-5} 
{CWE269+94}
 &  SSD & \underline{71.70\%} & \underline{75.99\%} & \underline{89.33\%}
 \tabularnewline
%\cline{2-5} \cline{3-5} \cline{4-5} \cline{5-5} 
{vs. CWE20}
 &  VulDeePecker & 79.26\%	& 73.30\% & 86.93\%
 \tabularnewline
%\cline{2-5} \cline{3-5} \cline{4-5} \cline{5-5} 
 &  CodeBERT &  85.47\% & 69.37\% & 84.56\%
 \tabularnewline
%\cline{2-5} \cline{3-5} \cline{4-5} \cline{5-5} 
 &  ReGVD &  90.40\%	& 62.13\% & 74.91\%
 \tabularnewline
\cline{2-5} \cline{3-5} \cline{4-5} \cline{5-5} 

& \multirow{2}{*}{\ourapp~(Ours)} & \textbf{63.26\%} & \textbf{82.44\%} & \textbf{91.74\%}\tabularnewline
%\cline{3-5} \cline{4-5} \cline{5-5} 
 &  & \textcolor{blue}{($\downarrow$ 8.44\%)} & \textcolor{blue}{($\uparrow$ 6.45\%)} & \textcolor{blue}{($\uparrow$ 2.41\%)}\tabularnewline
\hline 
%\hline 
 
% \multirow{7}{*}
% &  Standard DNN & 83.41\%	& 70.26\%	& 78.50\%
% \tabularnewline
%\cline{2-5} \cline{3-5} \cline{4-5} \cline{5-5} 
% &  Outlier Exposure &  86.30\%	& 67.15\%	& 77.00\%
% \tabularnewline
%\cline{2-5} \cline{3-5} \cline{4-5} \cline{5-5} 
%{CWE287-94}
% &  SSD & \underline{72.45\%}	& \underline{80.50\%}	& \underline{86.84\%}
% \tabularnewline
%\cline{2-5} \cline{3-5} \cline{4-5} \cline{5-5} 
%{vs. CWE416}
% &  VulDeePecker & 85.08\%	& 65.27\%	& 76.64\%
% \tabularnewline
%\cline{2-5} \cline{3-5} \cline{4-5} \cline{5-5} 
% &  CodeBERT & 84.60\%	& 63.47\%	& 74.56\%
% \tabularnewline
%\cline{2-5} \cline{3-5} \cline{4-5} \cline{5-5} 
% &  ReGVD  & 90.72\%	& 57.38\%	& 69.68\%
% \tabularnewline
%\cline{2-5} \cline{3-5} \cline{4-5} \cline{5-5} 

 %& \multirow{2}{*}{\ourapp~ (Ours)} & \textbf{60.88\%} & \textbf{81.00\%} & \textbf{88.12\%}\tabularnewline
%\cline{3-5} \cline{4-5} \cline{5-5} 
% &  & \textcolor{blue}{($\downarrow$ 11.57\%)} & \textcolor{blue}{($\uparrow$ 0.50\%)} & \textcolor{blue}{($\uparrow$ 1.28\%)}\tabularnewline
% \hline
 %\hline	
 
  \multirow{7}{*}
 &  Standard DNN & 80.44\%	& 73.22\%	& 91.77\%
 \tabularnewline
%\cline{2-5} \cline{3-5} \cline{4-5} \cline{5-5} 
 &  Outlier Exposure &  80.07\%	& 72.59\%	& 91.56\%
 \tabularnewline
%\cline{2-5} \cline{3-5} \cline{4-5} \cline{5-5} 
{CWE863+287}
 &  SSD & \underline{70.30\%}	& \underline{80.70\%}	& \underline{94.27\%}
 \tabularnewline
%\cline{2-5} \cline{3-5} \cline{4-5} \cline{5-5} 
vs. {CWE20}
 &  VulDeePecker & 78.37\%	& 75.01\%	& 92.39\%
 \tabularnewline
%\cline{2-5} \cline{3-5} \cline{4-5} \cline{5-5} 
 &  CodeBERT & 85.92\%	& 74.23\%	& 91.37\%
 \tabularnewline
%\cline{2-5} \cline{3-5} \cline{4-5} \cline{5-5} 
 &  ReGVD  & 89.40\%	& 62.29\%	& 83.87\%
 \tabularnewline
\cline{2-5} \cline{3-5} \cline{4-5} \cline{5-5} 

 & \multirow{2}{*}{\ourapp~(Ours)} & \textbf{47.85\%} & \textbf{88.76\%} & \textbf{97.16\%}\tabularnewline
%\cline{3-5} \cline{4-5} \cline{5-5} 
 &  & \textcolor{blue}{($\downarrow$ 22.45\%)} & \textcolor{blue}{($\uparrow$ 8.06\%)} & \textcolor{blue}{($\uparrow$ 2.89\%)}\tabularnewline
 \hline

 \multirow{7}{*}
 &  Standard DNN & 84.02\%	& 67.47\%	& 70.30\%
 \tabularnewline
%\cline{2-5} \cline{3-5} \cline{4-5} \cline{5-5} 
 &  Outlier Exposure &  87.82\%	& 66.34\%	& 69.24\%
 \tabularnewline
%\cline{2-5} \cline{3-5} \cline{4-5} \cline{5-5} 
{CWE287+862+94}
 &  SSD & \underline{71.08\%}	& \underline{74.95\%}	& \underline{78.97\%}
 \tabularnewline
%\cline{2-5} \cline{3-5} \cline{4-5} \cline{5-5} 
vs. {CWE416}
 &  VulDeePecker & 78.39\%	& 67.86\%	& 72.03\%
 \tabularnewline
%\cline{2-5} \cline{3-5} \cline{4-5} \cline{5-5} 
 &  CodeBERT & 91.62\%	& 68.64\%	& 67.69\%
 \tabularnewline
%\cline{2-5} \cline{3-5} \cline{4-5} \cline{5-5} 
 &  ReGVD  & 94.22\%	& 55.55\%	& 58.82\%
 \tabularnewline
\cline{2-5} \cline{3-5} \cline{4-5} \cline{5-5} 

 & \multirow{2}{*}{\ourapp~(Ours)} & \textbf{55.71\%} & \textbf{84.98\%} & \textbf{86.59\%}\tabularnewline
%\cline{3-5} \cline{4-5} \cline{5-5} 
 &  & \textcolor{blue}{($\downarrow$ 15.37\%)} & \textcolor{blue}{($\uparrow$ 10.03\%)} & \textcolor{blue}{($\uparrow$ 7.62\%)}\tabularnewline
 \hline
\end{tabular}}
\vspace{-5mm}
\end{table}

\revise{For this experiment, without loss of generality, we assume the presence of two and three different CWE categories within the in-distribution data, as shown in Table \ref{tab:my_label23t}. We randomly choose source code data from different CWE types to create the combinations to form the in-distribution data. Subsequently, for each case of the in-distribution data, after the training process, in the inference (testing) phase, we use source code data from other CWE categories as OOD data to evaluate the model's performance.}

\paragraph{\textbf{Results}}

\revise{The results in Table \ref{tab:my_label23t} once again demonstrate the effectiveness and superiority of our \ourapp~method compared to the baselines for OOD source code data identification in all various cases of the in-distribution data. Specifically, in these cases when there is more than one CWE category existing in the in-distribution data (e.g., two and three different CWE source code data categories), our  \ourapp~method still consistently obtains much higher performances on all pairs of the in-distribution and out-of-distribution source code data on the used measures. In these cases, on average, our \ourapp~method gains higher performances from around 14.22\%, 6.45\%, and 5.08\% than the baselines on the FPR (at TPR 95\%), AUROC, and AUPR measures, respectively.}

In Figure \ref{fig:averageresults}, we present the average values of the FPR, AUROC, and AUPR measures, respectively, for our \ourapp~method and baselines in all cases of the ID and OOD CWE categories mentioned in Tables \ref{tab:my_label1t} and \ref{tab:my_label23t}. These average results highlight the advancement of our \ourapp~method compared to the baselines. In short, on average, our \ourapp~method obtains a significantly higher performance from around 15.27\%, 7.39\%, and 4.93\% on the FPR, AUROC, and AUPR measures, respectively, in comparison with the baselines.

\vspace{3mm}
\colorbox{gray!20}{
\begin{minipage}{0.92\columnwidth}
\textbf{In conclusion for RQ2}: The experimental results in Table \ref{tab:my_label23t} again show the superiority of our proposed \ourapp~method over the baselines for OOD source code data identification on different cases regarding the various number of CWE categories existing in the in-distribution data. In concurrence with the results shown in Table \ref{tab:my_label1t}, our method demonstrates its noticeable advancement with much higher performances compared to the baselines on the used measures in almost all cases of the ID and OOD CWE source code data categories.
\end{minipage}
}
\vspace{1mm}

\vspace{1mm}
\textbf{RQ3: \rqthree}

\paragraph{\textbf{Approach}}

We compare the performance of our \ourapp~method to itself in the cases of using and without using cluster-contrastive learning mentioned in Eq. (\ref{eq:cluster_cl}) and the \ourmechanism~described in Eq. (\ref{eq:realx}) to investigate if these additional terms successfully boosting the code data representation learning facilitating out-of-distribution source code data identification. 

\vspace{0mm}
We experiment on some pairs of the in-distribution (ID) and out-of-distribution (OOD) CWE source code data categories as shown in Table \ref{tab:my_label2leo}. Note that in the case without using the cluster-contrastive learning and \ourmechanism, we denote the method as \ourapp-w/oCD.

\vspace{1mm}
\paragraph{\textbf{Results}}

\begin{table}[ht]
\vspace{0mm}
\caption{\revise{The results of our \ourapp~method for the FPR (at TPR 95\%), AUROC, and AUPR measures on the vulnerable source code of each OOD CWE category corresponding with specific ID data when using and without using the cluster-contrastive learning and the \ourmechanism. (The higher results in each pair of the ID and OOD data are in \textbf{bold}. The numbers highlighted in the blue color are the improvements of \ourapp~over \ourapp-w/oCD.)}}\label{tab:my_label2leo}
\centering{}
\vspace{0mm}
\resizebox{0.95\columnwidth}{!}{
% {|c|c|r|r|r|}
\begin{tabular}{ccrrr}
\hline 
%\textbf{ID and OOD} & \textbf{Methods} & \textbf{FPR} $\downarrow$ & \textbf{AUROC} $\uparrow$ & \textbf{AUPR} $\uparrow$
\textbf{ID and OOD} & \textbf{Methods} & \textbf{FPR} $\downarrow$ & \textbf{AUROC} $\uparrow$ & \textbf{AUPR} $\uparrow$
\tabularnewline
\hline 

\multirow{2}{*}
 &  \ourapp-woCD & 83.14\%	& 69.29\%	& 78.13\%
 \tabularnewline
\cline{2-5} \cline{3-5} \cline{4-5} \cline{5-5} {CWE269 vs. CWE200}

 & \multirow{2}{*}{\ourapp~}  &  \textbf{76.89\%}	& \textbf{79.33\%} & \textbf{85.32\%}\tabularnewline
 &  & \textcolor{blue}{($\downarrow$ 6.25\%)} & \textcolor{blue}{($\uparrow$ 10.04\%)} & \textcolor{blue}{($\uparrow$ 7.19\%)}\tabularnewline
 
\hline 
%\hline 

\multirow{2}{*}
 &  \ourapp-w/oCD &  78.07\%	& 73.54\% & 95.43\%
 \tabularnewline
\cline{2-5} \cline{3-5} \cline{4-5} \cline{5-5} {CWE94 vs. CWE20}

  & \multirow{2}{*}{\ourapp~}  & \textbf{51.70\%} & \textbf{84.37\%} & \textbf{97.72\%}\tabularnewline 
 &  & \textcolor{blue}{($\downarrow$ 26.37\%)} & \textcolor{blue}{($\uparrow$ 10.83\%)} & \textcolor{blue}{($\uparrow$ 2.29\%)}\tabularnewline
\hline 
%\hline 	

\multirow{2}{*}
& \ourapp-w/oCD & 81.13\%	& 62.99\%	& 84.96\%
\tabularnewline
\cline{2-5} \cline{3-5} \cline{4-5} \cline{5-5} {CWE287 vs. CWE416}

 & \multirow{2}{*}{\ourapp~}  & \textbf{69.71\%} & \textbf{76.75\%} & \textbf{90.57\%}\tabularnewline
 & & \textcolor{blue}{($\downarrow$ 11.42\%)} & \textcolor{blue}{($\uparrow$ 13.76\%)} & \textcolor{blue}{($\uparrow$ 5.61\%)}\tabularnewline
\hline
%\hline	

\multirow{2}{*}{CWE863+287}
 &  \ourapp-w/oCD &  77.33\%	&  74.23\%	&  92.27\%
 \tabularnewline
\cline{2-5} \cline{3-5} \cline{4-5} \cline{5-5}

 & \multirow{2}{*}{\ourapp~}  & \textbf{47.85\%} & \textbf{88.76\%} & \textbf{97.16\%}\tabularnewline {vs. CWE20}
 & & \textcolor{blue}{($\downarrow$ 29.48\%)} & \textcolor{blue}{($\uparrow$ 14.53\%)} & \textcolor{blue}{($\uparrow$ 4.89\%)}\tabularnewline
 \hline
 %\hline 

\multirow{2}{*}{CWE863+862}
&  \ourapp-w/oCD & 50.00\% & 79.15\%	& 62.60\%
\tabularnewline
\cline{2-5} \cline{3-5} \cline{4-5} \cline{5-5}

& \multirow{2}{*}{\ourapp~}  & \textbf{37.88\%} & \textbf{90.48\%} & \textbf{78.69\%}\tabularnewline {vs. CWE287}
 & & \textcolor{blue}{($\downarrow$ 12.12\%)} & \textcolor{blue}{($\uparrow$ 11.33\%)} & \textcolor{blue}{($\uparrow$ 16.09\%)}\tabularnewline
\hline

\multirow{2}{*}{CWE287+862+94}
&  \ourapp-w/oCD & 82.04\% & 64.39\%	& 69.96\%
\tabularnewline
\cline{2-5} \cline{3-5} \cline{4-5} \cline{5-5}

& \multirow{2}{*}{\ourapp~}  & \textbf{55.71\%} & \textbf{84.98\%} & \textbf{86.59\%}\tabularnewline {vs. CWE416}
 & & \textcolor{blue}{($\downarrow$ 26.33\%)} & \textcolor{blue}{($\uparrow$ 20.59\%)} & \textcolor{blue}{($\uparrow$ 16.63\%)}\tabularnewline
\hline

\end{tabular}}
\vspace{1mm}
\end{table}

The experimental results in Table \ref{tab:my_label2leo} show the effectiveness of cluster-contrastive learning and the \ourmechanism~in improving the source code data representation learning process to boost the model performance for OOD source code data identification. In particular, our method with the cluster-contrastive learning and \ourmechanism~terms (\textbf{\ourapp}) obtained considerably higher performances than itself without using these terms (denoted as \textbf{\ourapp-w/oCD}) in all cases of the in-distribution and out-of-distribution CWE categories.

\vspace{0mm}

\begin{figure}[ht]
\begin{centering}
\vspace{1mm}
\includegraphics[width=0.8\columnwidth]{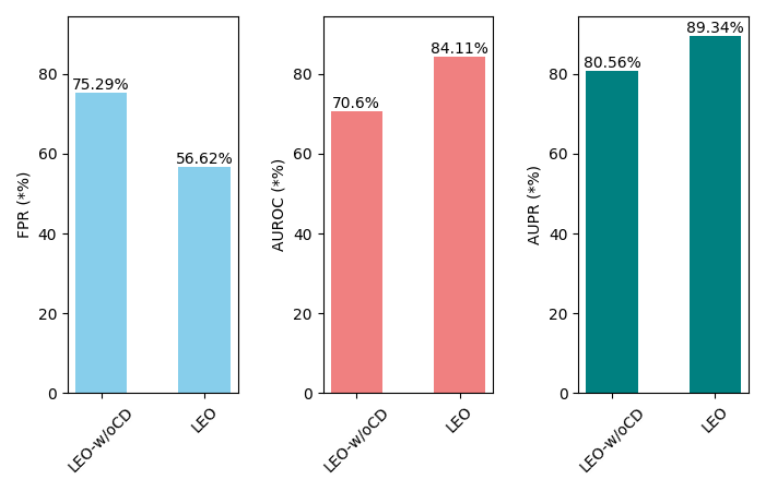}
\par\end{centering}\vspace{0mm}
\caption{\revise{The average results for the FPR (at TPR 95\%), AUROC, and AUPR measures, respectively, of \ourapp-w/oCD and \ourapp~in all cases of the in-distribution and out-of-distribution CWE source code data categories mentioned in Table \ref{tab:my_label2leo}.}}\label{fig:averageresults_leowwt}
\vspace{1mm}
\end{figure}%

The average experimental results for the FPR, AUROC, and AUPR measures of the \ourapp~and  \ourapp-w/oCD methods across all cases of the in-distribution (ID) and out-of-distribution (OOD) CWE source code data categories, as mentioned in Tables \ref{tab:my_label2leo}, are shown at Figure \ref{fig:averageresults_leowwt}. Particularly, on average, \ourapp~achieves much higher performances around 18.66\%, 13.51\%, and 8.78\%  on the FPR, AUROC, and AUPR measures, respectively, compared to \ourapp-w/oCD. These results underscore the substantial advancement of \ourapp~over \ourapp-w/oCD in all the used ID and OOD cases, as measured by the FPR, AUROC, and AUPR metrics.

\vspace{3mm}
\colorbox{gray!20}{
\begin{minipage}{0.9\columnwidth}
\textbf{In conclusion for RQ3}: The results mentioned in Table \ref{tab:my_label2leo} on the FPR (at TPR 95\%), AUROC, and AUPR measures show the benefits of using cluster-contrastive learning and the \ourmechanism~for boosting source code data representation learning. These terms help significantly improve the model performance for OOD source code data identification in all cases of the in-distribution and out-of-distribution CWE source code data categories.

\end{minipage}
}
\vspace{1mm}

\subsection{\textbf{Threats to validity }}

\subsubsection{\textbf{Construct validity}}
Key construct validity threats are if the assessments of our proposed \ourapp~method and baselines demonstrate their capability for out-of-distribution source code data identification. In the cyber security domain, the out-of-distribution source code data identification task helps determine if an input is from an in-distribution (ID) or out-of-distribution (OOD). That enables the model and security experts to take appropriate precautions and actions. In particular, this stage assists security analysts and software engineers in identifying potential out-of-distribution CWE source code data outside the established in-distribution CWE source code data categories before feeding input data into subsequent machine learning and deep learning modules, to harden the software security systems. To evaluate the performance of our \ourapp~method and baselines, we use three main measures widely used in out-of-distribution detection including FPR (at TPR 95\%), AUROC, and AUPR.

\vspace{1mm}
\subsubsection{\textbf{Internal validity}}
Key internal validity threats are relevant to the choice of hyper-parameter settings (i.e., optimizer, learning rate, number of layers in deep neural networks, etc.). It is worth noting that finding a set of optimal hyperparameter settings of deep neural networks is expensive due to a large number of trainable parameters. To train our method, we only use the common or default values for the hyperparameters. For example, we use the Adam optimizer; the learning rate equals $10^{-3}$; the size of neural layers is set in $\{100, 300\}$ while the scalar temperature $\tau$ used in the cluster-contrastive learning and the temperature $\nu$ for the Gumbel softmax distribution is set in $\{0.5, 1.0\}$. For the number of clusters used in cluster-contrastive learning aiming to improve the data representation learning process, we use it as a hyperparameter. In our paper, we detail the hyperparameter settings (i.e., the value-tuned ranges of the hyperparameters) in the released reproducible source code to support future replication studies.

\vspace{1mm}
\subsubsection{\textbf{External validity}}
Key external validity threats include whether our \ourapp~method will generalize across multiple CWE source code categories. We mitigated this problem by conducting the experiments on an up-to-date and big C/C++ dataset, \textit{namely DiverseVul}, \cite{chen2023diversevul}. The DiverseVul dataset contains 18,945 vulnerable functions and 330,492 non-vulnerable functions covering 150 CWEs. In our paper, we conduct the experiments on the top most dangerous and popular CWE categories including CWE-119 (Improper Restriction of Operations within the Bounds of a Memory Buffer), CWE-20 (Improper Input Validation), CWE-125 (Out-of-bounds Read), CWE-200 (Exposure of Sensitive Information to an Unauthorized Actor), CWE-787 (Out-of-bounds Write), CWE-416 (Use After Free), CWE-269 (Improper Privilege Management), CWE-94 (Improper Control of Generation of Code), CWE-190 (Integer Overflow or Wraparound), CWE-264 (Permissions, Privileges, and Access Controls), CWE-863 (Incorrect Authorization), CWE-862 (Missing Authorization), and CWE-287 (Improper Authentication).
\vspace{-1mm}
\section{Conclusion}\label{sec:conclusion}

In this paper, we have successfully proposed an innovative deep learning-based approach, a pioneering study, \revise{for out-of-distribution (OOD) source code data identification.} Our proposed \ourapp~method derived from the information theory, combined with the use of innovative cluster-contrastive learning and the \ourmechanism, has effectively learned the source code characteristics and figured out the semantic relationships inside and between source code data samples to boost the data representation learning facilitating OOD source code data identification. \revise{The extensive experimental results on the top most dangerous and popular CWE source code data categories demonstrate the effectiveness and advancement of our  \ourapp~method compared to the state-of-the-art baselines by a wide margin.}

\bibliographystyle{IEEEtran}
\bibliography{reference}

%\pagebreak
\newpage

\section{Appendix}\label{sec:appendix}

\subsection{\textbf{Additional experiments}}
Here we present some additional experiments of our \ourapp~method and baselines on other cases of the in-distribution (ID) and out-of-distribution (OOD) CWE source code data categories. The experimental results in Table \ref{tab:my_label_app} again show the effectiveness and superiority of our \ourapp~method compared to the baselines for OOD source code data identification by a wide margin. Notably, in these additional cases, on average, our LEO method obtains a significantly higher performance from around 10.59\%, 4.65\%, and 6.72\% on the FPR, AUROC, and AUPR measures, respectively, in comparison with the baseline approaches.

\begin{table}[ht]
\vspace{2mm}
\caption{\revise{The results of our \ourapp~method and baselines for the FPR (at TPR 95\%), AUROC, and AUPR measures on the vulnerable source code samples of each OOD CWE category corresponding with specific ID data. (The best results are in \textbf{bold} while the second highest results are in \underline{underline}. The numbers highlighted in blue represent the improvements of our \ourapp~method over the second-best baselines.)}}\label{tab:my_label_app}
\vspace{-1mm}
\centering{}
\resizebox{0.93\columnwidth}{!}{
\begin{tabular}{ccrrr}
\hline 
%\textbf{ID and OOD Datasets} & \textbf{Methods} & \textbf{FPR} $\downarrow$ & \textbf{AUROC} $\uparrow$ & \textbf{AUPR} $\uparrow$
\textbf{ID and OOD} & \textbf{Methods} & \textbf{FPR} $\downarrow$& \textbf{AUROC} $\uparrow$& \textbf{AUPR} $\uparrow$
\tabularnewline
\hline 

\multirow{7}{*}{CWE863 vs. CWE862}
 &  Standard DNN & 77.42\%	&  81.36\%	&  57.53\% 
 \tabularnewline
%\cline{2-5} \cline{3-5} \cline{4-5} \cline{5-5} 
 &  Outlier Exposure &  77.42\%	&  \underline{83.83\%}	&  62.19\%
 \tabularnewline
%\cline{2-5} \cline{3-5} \cline{4-5} \cline{5-5} 
 &  SSD & \underline{61.29\%}	&  75.95\%	&  \underline{66.83\%}
 \tabularnewline
%\cline{2-5} \cline{3-5} \cline{4-5} \cline{5-5} 
 &  VulDeePecker & 77.42\%	&  83.11\%	&  55.39\%
 \tabularnewline
%\cline{2-5} \cline{3-5} \cline{4-5} \cline{5-5} 
 &  CodeBERT  &  76.67\%	& 68.85\%	&  48.24\%
 \tabularnewline
%\cline{2-5} \cline{3-5} \cline{4-5} \cline{5-5} 
 &  ReGVD  &  96.67\%	&  66.95\%	&  37.56\%
 \tabularnewline
\cline{2-5} \cline{3-5} \cline{4-5} \cline{5-5} 

 & \multirow{2}{*}{\ourapp~(Ours)} & \textbf{54.84\%} & \textbf{84.31\%} & \textbf{70.76\%}\tabularnewline
%\cline{3-5} \cline{4-5} \cline{5-5} 
 &  & \textcolor{blue}{($\downarrow$ 6.45\%)} & \textcolor{blue}{($\uparrow$ 0.48\%)} & \textcolor{blue}{($\uparrow$ 3.93\%)}\tabularnewline
 \hline

\multirow{7}{*}{CWE190 vs. CWE787}
&  Standard DNN & 85.14\%	& 68.41\%	& 54.78\%
\tabularnewline
%\cline{2-5} \cline{3-5} \cline{4-5} \cline{5-5} 
&  Outlier Exposure &  \underline{83.36\%}	& 69.46\%	& 56.76\%
\tabularnewline
%\cline{2-5} \cline{3-5} \cline{4-5} \cline{5-5} 
&  SSD & 84.70\%	& \underline{71.02\%}	& \underline{57.76\%}
\tabularnewline
%\cline{2-5} \cline{3-5} \cline{4-5} \cline{5-5} 
&  VulDeePecker & 85.74\%	& 69.51\%	& 56.58\%
\tabularnewline
%\cline{2-5} \cline{3-5} \cline{4-5} \cline{5-5} 
&  CodeBERT  & 93.01\%	& 66.19\% & 47.95\%
\tabularnewline
%\cline{2-5} \cline{3-5} \cline{4-5} \cline{5-5} 
&  ReGVD  & 96.88\%	& 49.28\%	& 37.43\%
\tabularnewline
\cline{2-5} \cline{3-5} \cline{4-5} \cline{5-5} 

& \multirow{2}{*}{\ourapp~(Ours)} & \textbf{66.42\%} & \textbf{80.39\%} & \textbf{71.58\%}\tabularnewline
%\cline{3-5} \cline{4-5} \cline{5-5} 
 &  & \textcolor{blue}{($\downarrow$ 16.94\%)} & \textcolor{blue}{($\uparrow$ 9.37\%)} & \textcolor{blue}{($\uparrow$ 13.82\%)}\tabularnewline	
 
 \hline
 
\multirow{7}{*}{CWE94 vs. CWE269}
&  Standard DNN & 83.33\%	& 77.51\%	& 63.73\%
\tabularnewline
%\cline{2-5} \cline{3-5} \cline{4-5} \cline{5-5} 
&  Outlier Exposure &  83.33\%	& 71.61\%	& 56.74\%
\tabularnewline
%\cline{2-5} \cline{3-5} \cline{4-5} \cline{5-5} 
&  SSD & \underline{66.67\%}	& \underline{79.70\%}	& \underline{68.64\%}
\tabularnewline
%\cline{2-5} \cline{3-5} \cline{4-5} \cline{5-5} 
&  VulDeePecker & 80.00\%	& 70.25\%	& 57.17\%
\tabularnewline
%\cline{2-5} \cline{3-5} \cline{4-5} \cline{5-5} 
&  CodeBERT  & 89.89\%	& 62.70\%	& 45.01\%
\tabularnewline
%\cline{2-5} \cline{3-5} \cline{4-5} \cline{5-5} 
&  ReGVD  & 93.26\%	& 63.49\%	& 44.22\%
\tabularnewline
\cline{2-5} \cline{3-5} \cline{4-5} \cline{5-5} 
&  \multirow{2}{*}{\ourapp~(Ours)}  & \textbf{61.11\%} & \textbf{79.88\%} & \textbf{70.57\%}
\tabularnewline
%\cline{3-5} \cline{4-5} \cline{5-5} 
 &  & \textcolor{blue}{($\downarrow$ 5.56\%)} & \textcolor{blue}{($\uparrow$ 0.18\%)} & \textcolor{blue}{($\uparrow$ 1.93\%)}\tabularnewline
\hline
 
%\hline
\multirow{7}{*}{CWE264 vs. CWE200}
&  Standard DNN & 80.49\%	& 71.59\%	& 64.71\%
\tabularnewline
%\cline{2-5} \cline{3-5} \cline{4-5} \cline{5-5} 
&  Outlier Exposure &  83.71\% &  72.13\% & 65.96\%
\tabularnewline
%\cline{2-5} \cline{3-5} \cline{4-5} \cline{5-5} 
&  SSD & \underline{72.54\%}	& \underline{77.04\%}	& \underline{73.67\%}
\tabularnewline
%\cline{2-5} \cline{3-5} \cline{4-5} \cline{5-5} 
&  VulDeePecker & 78.79\%	& 72.24\%	& 67.30\%
\tabularnewline
%\cline{2-5} \cline{3-5} \cline{4-5} \cline{5-5} 
&  CodeBERT  & 90.32\%	& 70.02\%	& 59.78\%
\tabularnewline
%\cline{2-5} \cline{3-5} \cline{4-5} \cline{5-5} 
&  ReGVD  & 91.27\%	& 57.50\%	& 53.21\%
\tabularnewline
\cline{2-5} \cline{3-5} \cline{4-5} \cline{5-5} 
&  \multirow{2}{*}{\ourapp~(Ours)}  & \textbf{64.96\%} & \textbf{81.25\%} & \textbf{76.69\%}
\tabularnewline
%\cline{3-5} \cline{4-5} \cline{5-5} 
 &  & \textcolor{blue}{($\downarrow$ 7.58\%)} & \textcolor{blue}{($\uparrow$ 4.21\%)} & \textcolor{blue}{($\uparrow$ 3.02\%)}\tabularnewline
\hline
%\hline	
 
 \multirow{7}{*}
 &  Standard DNN & 83.44\%	& 70.97\%	& 57.45\%
 \tabularnewline
%\cline{2-5} \cline{3-5} \cline{4-5} \cline{5-5} 
 &  Outlier Exposure &  83.68\%	& 69.95\%	& 56.81\%
 \tabularnewline
%\cline{2-5} \cline{3-5} \cline{4-5} \cline{5-5} 
{CWE190+287}
 &  SSD & \underline{81.83\%}	& \underline{72.19\%}	& \underline{60.47\%}
 \tabularnewline
%\cline{2-5} \cline{3-5} \cline{4-5} \cline{5-5} 
vs. {CWE119}
 &  VulDeePecker & 82.94\%	& 70.73\%	& 56.96\%
 \tabularnewline
%\cline{2-5} \cline{3-5} \cline{4-5} \cline{5-5} 
 &  CodeBERT & 87.75\%	& 65.71\%	& 51.50\%
 \tabularnewline
%\cline{2-5} \cline{3-5} \cline{4-5} \cline{5-5} 
 &  ReGVD  & 97.03\%	& 48.27\%	& 35.76\%
 \tabularnewline
\cline{2-5} \cline{3-5} \cline{4-5} \cline{5-5} 
 %&  \ourapp~ (Ours)  & \textbf{0.26\%} & \textbf{99.75\%} & \textbf{92.44\%}
 %\tabularnewline
 & \multirow{2}{*}{\ourapp~(Ours)} & \textbf{71.32\%} & \textbf{75.76\%} & \textbf{67.59\%}\tabularnewline
%\cline{3-5} \cline{4-5} \cline{5-5} 
 &  & \textcolor{blue}{($\downarrow$ 10.51\%)} & \textcolor{blue}{($\uparrow$ 3.57\%)} & \textcolor{blue}{($\uparrow$ 7.12\%)}\tabularnewline
 \hline

  \multirow{7}{*}
 &  Standard DNN & 62.12\%	& 81.80\%	& 47.14\%
 \tabularnewline
%\cline{2-5} \cline{3-5} \cline{4-5} \cline{5-5} 
 &  Outlier Exposure &  59.09\%	& 82.28\%	& 46.47\%
 \tabularnewline
%\cline{2-5} \cline{3-5} \cline{4-5} \cline{5-5} 
{CWE863+862+94}
 &  SSD & \underline{51.52\%}	& \underline{86.61\%}	& \underline{57.95\%}
 \tabularnewline
%\cline{2-5} \cline{3-5} \cline{4-5} \cline{5-5} 
vs. {CWE287}
 &  VulDeePecker & 54.55\%	& 77.60\%	& 48.09\%
 \tabularnewline
%\cline{2-5} \cline{3-5} \cline{4-5} \cline{5-5} 
 &  CodeBERT & 86.15\%	& 61.30\%	& 18.89\%
 \tabularnewline
%\cline{2-5} \cline{3-5} \cline{4-5} \cline{5-5} 
 &  ReGVD  & 87.88\%	& 51.36\%	& 17.04\%
 \tabularnewline
\cline{2-5} \cline{3-5} \cline{4-5} \cline{5-5} 
 %&  \ourapp~ (Ours)  & \textbf{0.26\%} & \textbf{99.75\%} & \textbf{92.44\%}
 %\tabularnewline
 & \multirow{2}{*}{\ourapp~(Ours)} & \textbf{39.39\%} & \textbf{90.64\%} & \textbf{70.85\%}\tabularnewline
%\cline{3-5} \cline{4-5} \cline{5-5} 
 &  & \textcolor{blue}{($\downarrow$ 12.13\%)} & \textcolor{blue}{($\uparrow$ 4.03\%)} & \textcolor{blue}{($\uparrow$ 12.9\%)}\tabularnewline
 \hline
 
   \multirow{7}{*}
 &  Standard DNN & 85.23\%	& 71.03\%	& 64.04\%
 \tabularnewline
%\cline{2-5} \cline{3-5} \cline{4-5} \cline{5-5} 
 &  Outlier Exposure &  85.80\%	& 70.61\%	& 63.69\%
 \tabularnewline
%\cline{2-5} \cline{3-5} \cline{4-5} \cline{5-5} 
{CWE269+862+94}
 &  SSD & \underline{67.99\%}	& \underline{81.90\%}	& \underline{78.56\%}
 \tabularnewline
%\cline{2-5} \cline{3-5} \cline{4-5} \cline{5-5} 
vs. {CWE200}
 &  VulDeePecker & 85.61\%	& 70.66\%	& 63.13\%
 \tabularnewline
%\cline{2-5} \cline{3-5} \cline{4-5} \cline{5-5} 
 &  CodeBERT & 82.35\%	& 74.90\%	& 67.11\%
 \tabularnewline
%\cline{2-5} \cline{3-5} \cline{4-5} \cline{5-5} 
 &  ReGVD  & 84.47\%	& 64.56\%	& 58.98\%
 \tabularnewline
\cline{2-5} \cline{3-5} \cline{4-5} \cline{5-5} 
 %&  \ourapp~ (Ours)  & \textbf{0.26\%} & \textbf{99.75\%} & \textbf{92.44\%}
 %\tabularnewline
 & \multirow{2}{*}{\ourapp~(Ours)} & \textbf{54.36\%} & \textbf{84.71\%} & \textbf{82.89\%}\tabularnewline
%\cline{3-5} \cline{4-5} \cline{5-5} 
 &  & \textcolor{blue}{($\downarrow$ 13.63\%)} & \textcolor{blue}{($\uparrow$ 2.81\%)} & \textcolor{blue}{($\uparrow$ 4.33\%)}\tabularnewline
 \hline
 
\end{tabular}}
\vspace{0mm}
\end{table}
\end{document}